\documentclass[reprint,superscriptaddress,amsmath,amssymb,aps,pre]{revtex4-2}
\pdfoutput=1
\usepackage{graphicx}
\usepackage{dcolumn}
\usepackage{bm}
\usepackage{hyperref}
\usepackage{units}
\usepackage[english]{babel}

\usepackage{mathtools}
\usepackage{dsfont}

\begin{document}

\title{Human Learning of Hierarchical Graphs}

\author{Xiaohuan Xia}
\author{Andrei A. Klishin}
\affiliation{
Department of Bioengineering, University of Pennsylvania, Philadelphia, PA 19104 USA
}
\author{Jennifer Stiso}
\affiliation{
Department of Bioengineering, University of Pennsylvania, Philadelphia, PA 19104 USA
}
\author{Christopher W. Lynn}
\affiliation{Joseph Henry Laboratories of Physics, Princeton University, Princeton, NJ 08544, USA
}
\affiliation{Initiative for the Theoretical Sciences, Graduate Center, City University of New York, New York, NY 10016, USA
}
\author{Ari E. Kahn}
\affiliation{Princeton Neuroscience Institute, Princeton University, Princeton, NJ 08544 USA
}
\author{Lorenzo Caciagli}
\affiliation{
Department of Bioengineering, University of Pennsylvania, Philadelphia, PA 19104 USA
}
\author{Dani S. Bassett}
\email{dsb@seas.upenn.edu}
\affiliation{
Department of Bioengineering, University of Pennsylvania, Philadelphia, PA 19104 USA
}
\affiliation{
Department of Physics and Astronomy, University of Pennsylvania, Philadelphia, PA 19104 USA
}
\affiliation{
Department of Electrical \& Systems Engineering, University of Pennsylvania, Philadelphia, PA 19104 USA
}
\affiliation{
Department of Neurology, University of Pennsylvania, Philadelphia, PA 19104 USA
}
\affiliation{
Department of Psychiatry, University of Pennsylvania, Philadelphia, PA 19104 USA
}
\affiliation{
Santa Fe Institute, Santa Fe, NM 87501 USA
}

\date{August 28, 2023}

\begin{abstract}
Humans are constantly exposed to sequences of events in the environment. Those sequences frequently evince statistical regularities, such as the probabilities with which one event transitions to another. Collectively, inter-event transition probabilities can be modeled as a graph or network.
Many real-world networks are organized hierarchically and understanding how these networks are learned by humans is an ongoing aim of current investigations.
While much is known about how humans learn basic transition graph topology, whether and to what degree humans can learn hierarchical structures in such graphs remains unknown.
Here, we investigate how humans learn hierarchical graphs of the Sierpiński family using computer simulations and behavioral laboratory experiments.
We probe the mental estimates of transition probabilities via the \emph{surprisal effect}: a phenomenon in which humans react more slowly to less expected transitions, such as those between communities or modules in the network.
Using mean-field predictions and numerical simulations, we show that surprisal effects are stronger for finer-level than coarser-level hierarchical transitions.
Notably, surprisal effects at coarser levels of the hierarchy are difficult to detect for limited learning times or in small samples.
Using a serial response experiment with human participants (n=$100$), we replicate our predictions by detecting a surprisal effect at the finer-level of the hierarchy but not at the coarser-level of the hierarchy. To further explain our findings, we evaluate the presence of a trade-off in learning, whereby humans who learned the finer-level of the hierarchy better tended to learn the coarser-level worse, and \emph{vice versa}.
Taken together, our computational and experimental studies elucidate the processes by which humans learn sequential events in hierarchical contexts.
More broadly, our work charts a road map for future investigation of the neural underpinnings and behavioral manifestations of graph learning.
\end{abstract}

\maketitle

\begin{figure*}[t]
    \centering
	\includegraphics[width=\textwidth, height=4.5cm]{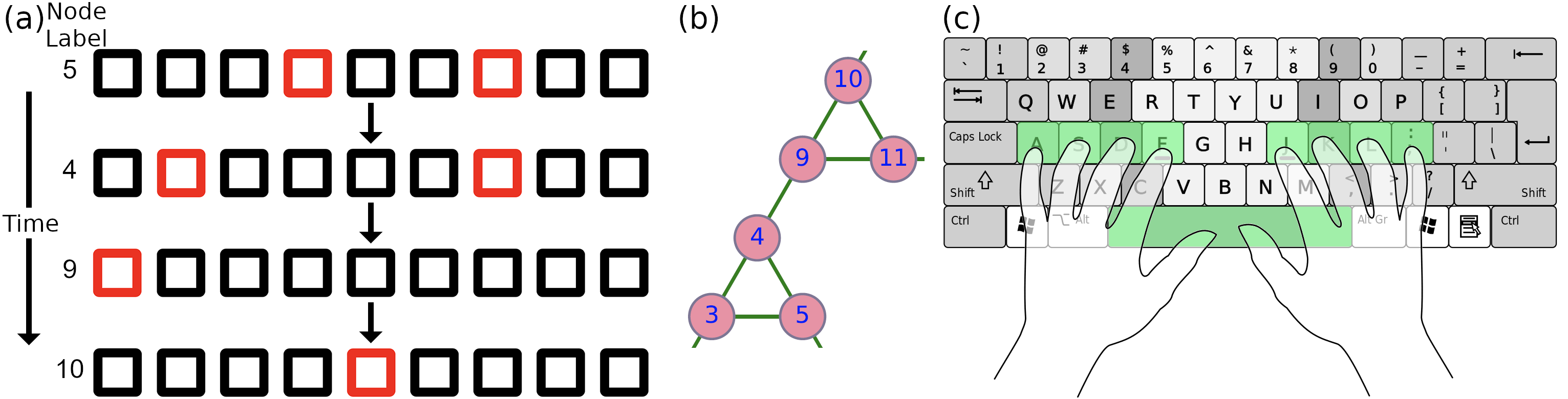}
	\caption{\textbf{Schematic of the task design.}
	\emph{(a)} Example sequence of visual stimuli; each row is shown to the participant one at a time.
	In this example, each row represents a unique color pattern of nine squares which corresponds to a unique node in a transition graph.
	For each participant, a sequence of $1500$ stimuli was drawn via a random walk on the same three-level Sierpiński graph ${}^{3}S_p^n$ (Methods).
	\emph{(b)} A part of the transition graph that involves the nodes in panel (a).
	The mapping between the color pattern in panel (a) and the node index in panel (b) was shuffled uniformly at random across participants so that any systematic biases of reactions to certain color patterns would be balanced across nodes.
	\emph{(c)} Hand placement; each of the keys highlighted in green corresponds to a square in any row of panel (a).
	When the squares were highlighted in red in panel (a), the participants were asked to press the corresponding keyboard input combinations, which were drawn from a total of $27$ possible combinations that did not require coordination between the two hands.}
	\label{Fig0.schematic}
\end{figure*}

\section{Introduction}
Humans perceive the world around them as a temporal sequence of consecutive events.
Such a sequence can be characterized by transition rules that specify which event is followed by which other events.
Transition rules are probabilistic; given a history of events, there are multiple candidate subsequent events, and each candidate is associated with a given transition probability.
The process whereby humans perceive and encode these transitions is called \textit{statistical learning} \cite{SLInfant,SLVisual,SLAuditory}, and manifests in many human activities. Some examples include learning visual patterns \cite{SLVisual,SLVisual2,SLVisual3,ProcessStructure,ProbMotorSeqLearnability}
or auditory sequences \cite{SLAuditory}, acquiring a first language \cite{AcqLang}, learning abstract relationships between objects \cite{Garvert2017, Constantinescu2016},
and understanding the structure of social networks \cite{IdSocial}.
Statistical learning can be studied by modeling events and their transitions using a transition \textit{graph} or \textit{network}: a mathematical object composed of nodes and edges that connect nodes.
In such a transition graph, a node represents a unique event, and a weighted edge between two nodes $a$ and $b$ represents the probability that event $b$ follows or precedes event $a$.
Sequences can then be generated by following transition rules defined by the transition graph.
In experiments, as humans perceive these sequences, they react to different transitions with different amounts of time, which reveals how the transition relations---and thus the underlying statistical structure---are learned \cite{ProcessStructure,ProbMotorSeqLearnability}.

Recent literature in statistical learning supports the view that humans are sensitive to different graph structures underlying transition probabilities \cite{STATStructure1,STATStructure2,SLInfant,SLVisual2,Lynn_learn_net2020}.
For example, when displaying action cues drawn from transition graphs, humans can detect differences in individual transition probabilities.
Specifically, they react slower to transitions with a lower probability and faster to transitions with a higher probability \cite{ProbMotorSeqLearnability,Lynn_info_process2020}.
This reaction time slowing is sometimes referred to as a ``surprisal effect."
Surprisal effects are also observed in response to mesoscale and global statistics of transition graphs. Specifically, humans react more quickly to cues drawn from a graph with a community structure than a graph without a community structure \cite{ProcessStructure,ProbMotorSeqLearnability,MentalErrors}. Further, humans react more slowly to individual transitions connecting different communities than to transitions within communities, even when the transition probabilities themselves are identical \cite{ProbMotorSeqLearnability}.
Additionally, reaction times in response to previously unseen transitions vary in proportion to the topological distance between nodes \cite{MentalErrors}.
Together, these findings indicate that humans develop a mental representation of transition graphs that differs from the true transition structure, resulting in different reaction times to elements with the same transition probability \cite{MentalErrors}.

Notably, maximum entropy models of the statistical learning process predict the above observations in human behavior \cite{MentalErrors}. The key parameter in such models, $\beta$, controls the rate of errors in memory when updating the mental model of the transition graph \cite{MentalErrors}.
These memory errors can lead to mental models that solve problems about the transition structure more accurately and flexibly than models without errors \cite{Momennejad2017}, because the memory errors intrinsically capture mesoscale information that is not evident in the one-step transition matrix.
When considering how to expand such models to real-world systems, it becomes important to acknowledge that many real transition structures are hierarchically organized across more than two levels. Examples include Wikipedia networks \cite{Hier_wiki}, email networks \cite{Hier_EmailNet}, the World Wide Web, and the Semantic Web \cite{Hier_MoreNets}.
Building accurate mental representations of these hierarchical structures is crucial for human problem solving \cite{Botvinick2009} and is evident from human behavior \cite{Eckstein2020}.
Some studies have examined the learning of rules at different abstraction or hierarchical levels in a task \cite{task-rule-hierarchy_2013},
and a recent work employing a classification task with two levels of abstraction has shown that humans can learn hierarchical organization in feature-based categorization tasks \cite{HierConceptLearn}.
Yet, it remains unclear how well the maximum entropy model \cite{MentalErrors} can effectively capture the learning of hierarchy in transition probabilities.
One main challenge to modeling this scenario is to define a relatively small, simple, and multilevel hierarchical graph that humans could feasibly learn within an experimental session.

In this study, we aimed to characterize processes that underlie the learning of hierarchy in graphs which encode transition probabilities between stimuli.
We modeled sequences of stimuli as unbiased random walks on the graph and used such sequences in the experiment (Fig. \ref{Fig0.schematic}).
We then identified several theoretical and practical properties for a candidate graph to model stimulus transitions:
(1) hierarchical community structure with at least three well-defined hierarchical levels, allowing us to extend our prior study of modular graphs \cite{MentalErrors};
(2) symmetric transition probabilities such that the probability of transitioning from $a$ to $b$ is the same as the probability of transitioning from $b$ to $a$;
(3) equal transition probabilities between all connected nodes; and
(4) small graph size, so that humans can learn the graph structure during a single experimental session.
With these considerations in mind, we selected the Sierpiński graph family \cite{Survey}, which provides a natural definition of hierarchical levels on nodes and edges \cite{Survey}.
Specifically, we chose a three-level regularized Sierpiński graph (Fig. \ref{Fig2.CCS_p}(a) Left panel) to address the question of whether the maximum entropy models of statistical learning can capture the surprisal effect beyond the first two hierarchical levels of the graph.

Prior work indicates that when humans learn modular transition graphs, they are more surprised at the transitions connecting modules than at the transitions within modules \cite{ProcessStructure}.
This difference in surprisal represents a two-level hierarchical effect: Humans react faster during transitions at the first level (finer level) than during transitions at the second level (coarser level) \cite{ProcessStructure}.
A natural aim is to generalize this two-level modular surprisal effect to a more general hierarchical graph.
In such a generalization, we hypothesize that humans will react faster during transitions that occur at a given hierarchical level than during transitions that occur at a coarser hierarchical level.
To test this hypothesis, we devise stochastic computer simulations and leverage data from a human experiment.
We then use the maximum entropy model of human perception to predict the surprisal effect on several different Sierpiński graphs; we do so first analytically in an infinite time horizon, and then in a finite time horizon using stochastic simulations.
To validate our predictions, we run a statistical learning experiment that features a walk of $1500$ steps---consistent with previous work \cite{MentalErrors}---on the transition graph ${}^{3}S_3^3$.
We then fit both the maximum entropy model and a linear mixed effects model to the collected empirical data to test our hypothesis at both the first and the second levels of the graph's hierarchy.

Collectively, our results show that human learners respond to a graph's coarser-scale structure more slowly than to a graph's finer-scale structure. Further, our findings indicate that to detect learning on coarser-scale structures, an experimenter might need to collect significantly more samples than they would to detect learning on finer-scale structures. Notably, we also observed a striking negative across-subjects correlation between the surprisal effect at the coarser scale and the surprisal effect at the finer scale, indicating that participants who learned the coarser-scale structures well tended to learn the fine-scale poorly, and \emph{vice versa}. 
This result interestingly suggests the existence of a trade-off in learning, whereby participants learn one hierarchical level at the expense of learning the other hierarchical level. Taken together, our results comprise a body of work that serves as a starting point in the investigation of how humans learn hierarchical graphs.

\section{Methods}

\subsection{Experimental Setup for the Probabilistic Sequential Motor Learning Task}

In this study, we used a serial response task to probe how humans learn hierarchical transition structure from a sequence. During this task running on Amazon's Mechanical Turk platform, human participants were shown a sequence of stimuli and asked to respond to each stimulus.
The probability of a transition between any two consecutive stimuli was governed by a transition graph.
To perform the task, human learners were asked to respond to each transition as quickly and as accurately as possible. The presentation of stimuli was self-paced, and the next stimulus was not displayed until a correct response was given to the previous stimulus. Participant reaction times were then a proxy for learning; swifter reaction times indicated better learning than slower reaction times. This study was approved by the Institutional Review Board of the University of Pennsylvania.
Written informed consent was obtained from all participants, in accordance with the Declaration of Helsinki.

In our experimental paradigm adapted from \cite{ProbMotorSeqLearnability}, the participants were instructed to respond quickly and accurately to a sequence of stimuli in a probabilistic sequential motor task (Fig. \ref{Fig0.schematic}).
During the task, each stimulus was a horizontal row of nine squares, with a unique combination of squares highlighted in red. Each square in the stimulus mapped to a key on the keyboard. Participants were told that their goal was to press the keys that were highlighted in red as quickly and accurately as possible.
The squares from left to right corresponded to keys `a', `s', `d', `f', `space', `j', `k', `l', and `;'.
The first four keys corresponded to the four fingers on the left hand starting from the fifth finger;
the last four keys corresponded to the four fingers on the right hand, starting from the index finger;
and the `space' key corresponded to the thumb, with participants being given the option to choose whichever thumb (left or right) suited them best. Participants were instructed to keep their hands in a fixed position in order speed up their responses.
If a participant's keypress was correct, there would be a delay of $50$ ms before the pattern of squares changed to the next one.
If a participant's keypress was incorrect, the message `Error!' would be displayed below the stimulus, and would remain on the screen until the participant pressed the correct key(s).
If there was no response for over one minute, the experiment would end.

Each stimulus corresponded to $1$ of $27$ unique keyboard input combinations. The combination comprised either a single key or two keys. Single keys could be any key listed above except `space'. Two-key combinations were either two keys from the same hand, or one key from one hand and the other key from the thumb. This configuration gives $28$ unique keyboard input combinations. The graphs used to generate sequences had only $27$ nodes; accordingly, for each participant, one keyboard input combination was not seen. We randomly excluded either `d'+`space' or `k'+`space' (but not both) in all participants.
These specific combinations were chosen because the right middle finger and thumb combination (`k'+`space' in our setup) yielded the slowest average reaction time in prior work \cite{ProbMotorSeqLearnability}, and we assumed a similar phenomenon to apply to the left hand. Thus, each unique stimulus had a unique combination of red/grey outlines, which mapped to a unique key press and unique node in the underlying graph (Fig. \ref{Fig0.schematic}). The keyboard input combinations were assigned to nodes in the ground truth graph at random across all participants.

The sequence of stimuli each participant saw was determined by a walk on a Sierpinski triangle graph (see later section on Ground Truth Graph Construction). Each walk was made from a combination of random and Hamiltonian walks on the graph \cite{SLVisual2}; the latter being commonly included in tasks of this kind to allow an assessment of the effects of recency \cite{SLVisual2,ProbMotorSeqLearnability}.
Specifically, the first $700$ steps of the $1500$-step sequence were drawn from random walks;
the subsequent $800$ steps comprised eight $100$-step sequences, each of which comprised a random walk sequence of length $73$ followed by a Hamiltonian walk sequence of length $27$.
The number $27$ was to ensure that the Hamiltonian walk covered all nodes.

We collected $208$ participants, but we excluded the first $10$ participants because the experiment was a preliminary test run, and then we excluded $98$ participants for not completing the task, leaving $100$ participants. All analyses were done on these $100$ participants after the exclusion criteria.
As per the experiment instruction, participants were financially compensated only if they completed the entire experiment. But after receiving some data, we relaxed our compensation policy to remunerate a base \$10 amount to any participant who finished at least $300$ steps of the graph learning section of the experiment, or who after the experiment emailed us confirming that they had made a reasonable attempt and hence to voice a complaint regarding compensation. In addition, if the participant completed the task with a performance accuracy that was $\geq90\%$, they would receive a bonus of \$1.5.
To ensure performance quality, we included a quiz before the task to test participant's understanding of the task, and they had to pass the quiz in order to proceed to the task.
One participant did not disclose their age; the remaining $99$ participants' ages have a mean of $37.4$, a standard deviation of $9.5$, a minimum of $22$, and a maximum of $64$. Most participants reported their sex assigned at birth as male ($37$/$63$ female/male) and their gender as man ($38$/$62$ women/men). The reported race/ethnicity of the participants was as follows: $11$ were African-American/Black; $3$ were Asian/Asian-American; $5$ were Hispanic/Latino; and $81$ were White.
Four individuals were left-handed, and $96$ were right-handed.

\begin{figure*}[t]
    \centering
	\includegraphics[width=\textwidth, height=9.5cm]{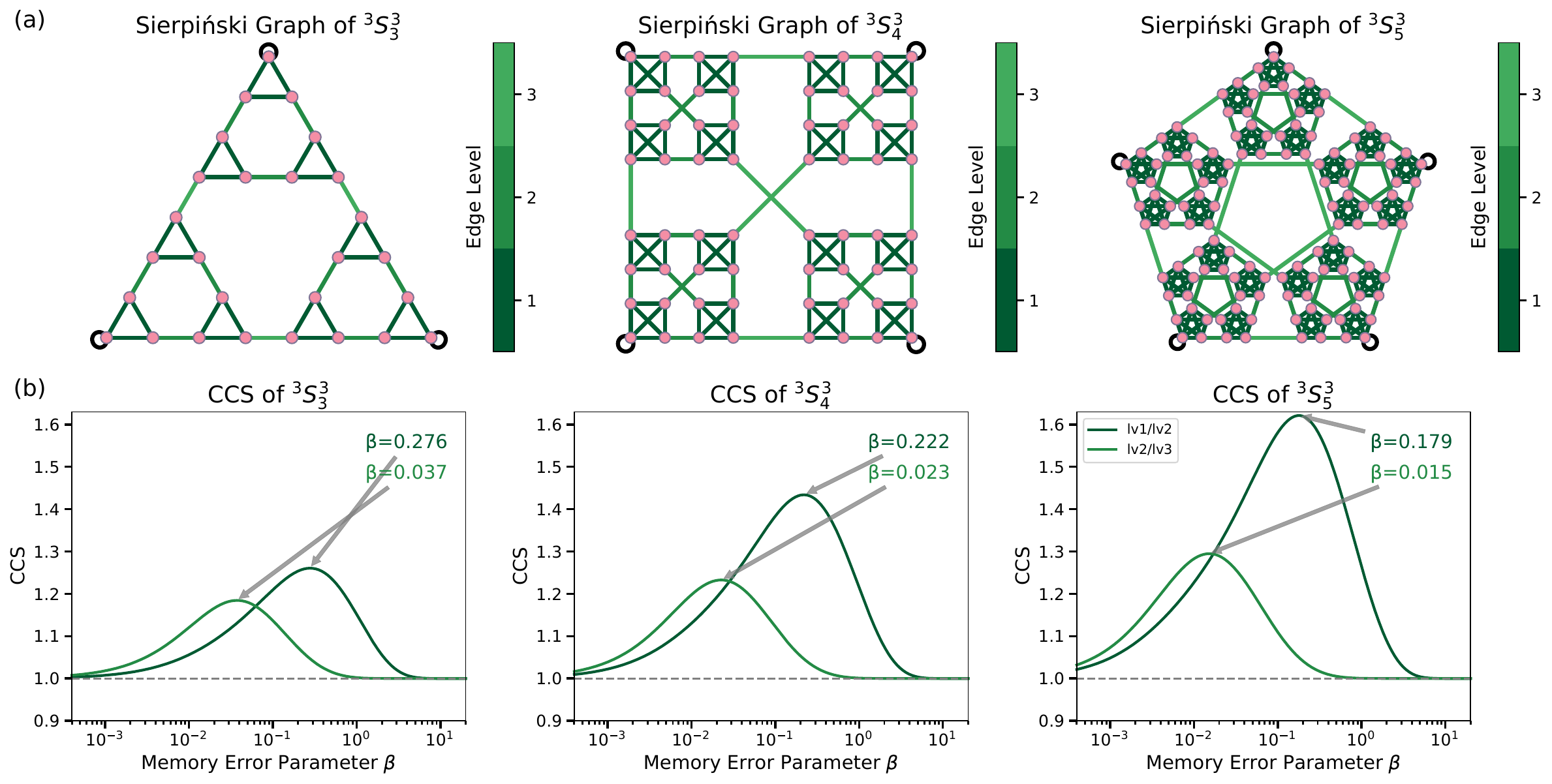}
	\caption{\textbf{Predictions of human learning and its dependence on hierarchy in the structure of transition probabilities between stimuli.} Here we use a validated model of human perception to predict how humans will respond to sequential information drawn from a graph topology \cite{MentalErrors}. A key indicator of human learning is a slowing of reaction time at the boundary between clusters in the graph. This slowing is referred to as the cross-cluster surprisal (CCS), which we show here for self-loop regularized level-$3$ Sierpiński graphs with different bases.
	\emph{(a)} Visualizations of Sierpiński graphs of base three, four, and five with a power of three. Nodes are shown in pink and edges are shown in green, except for the self-loop edges that are shown in black, because they do not belong to any well-defined edge level. The saturation of the green indicates the level of the hierarchy at which the edge is defined; we refer to this level as the edge level in the color bar label. We use a bottom-up convention for levels, meaning that the finest level is level-$1$ and the level value increases as the scale increases.
	\emph{(b)} The cross-cluster surprisal (CCS) for the corresponding Sierpiński graphs in panel (a) as a function of $\beta$: the rate of error in memory when updating the mental model of the transition graph. The $\beta$ value at which the cross-cluster surprisal peaks is marked for both levels of the graph's hierarchy.}
	\label{Fig2.CCS_p}
\end{figure*}

\begin{figure*}[t]
    \centering
	\includegraphics[width=\textwidth, height=9.5cm]{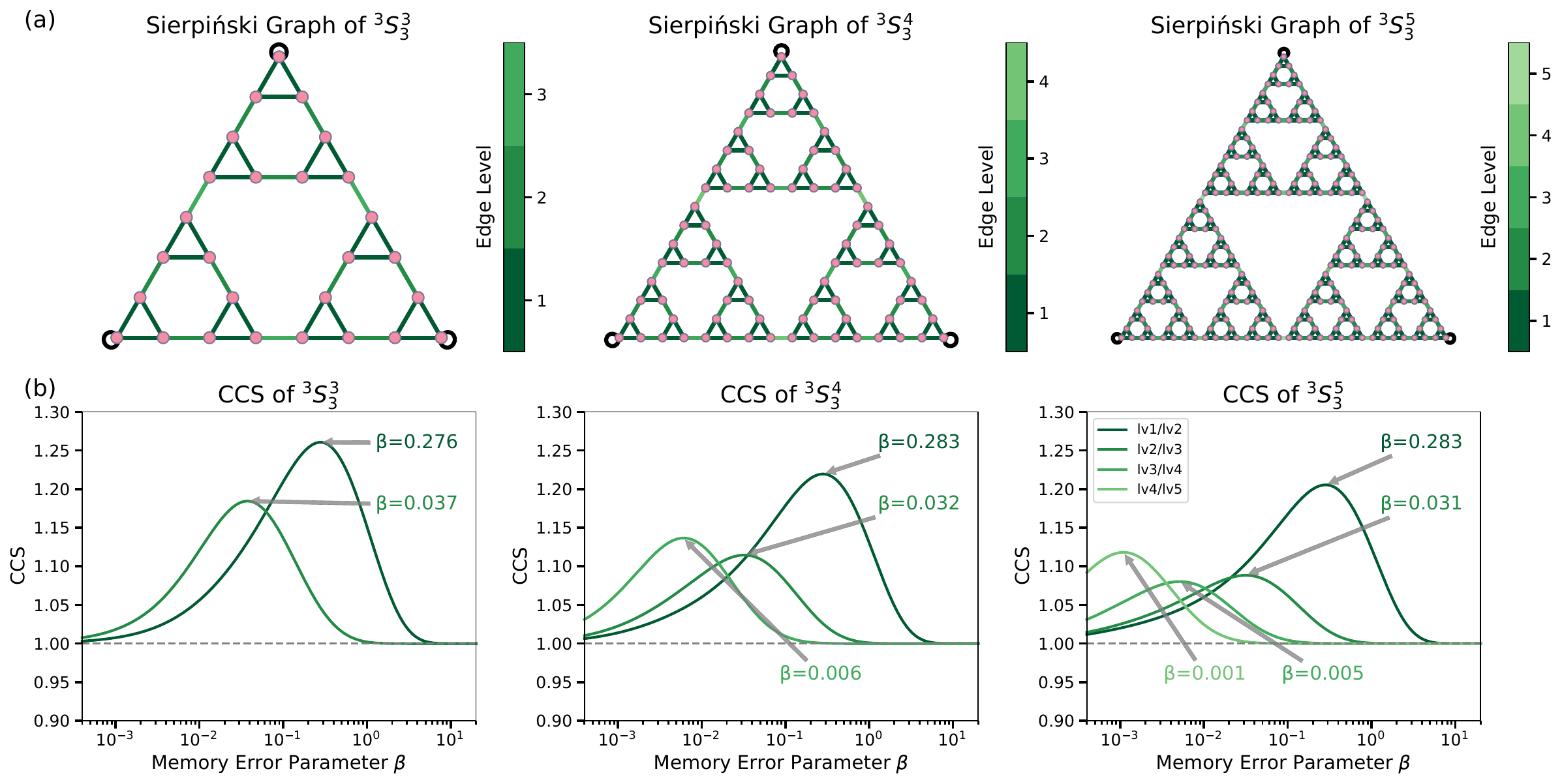}
	\caption{\textbf{Predictions of human learning and its dependence on hierarchy in self-loop regularized base-3 Sierpiński graphs that encode transition probabilities between stimuli.} Here again we use a validated model of human perception to predict how humans will respond to sequential information drawn from a graph topology \cite{MentalErrors}.
	\emph{(a)} Visualizations of Sierpiński graphs of power three, four, and five with a base of three. Nodes are shown in pink and edges are shown in green, except for the self-loop edges that are shown in black, because they do not belong to any well-defined edge level. The saturation of the green indicates the level of the hierarchy at which the edge exists; we refer to this level as the edge level in the color bar label.
	\emph{(b)} The cross-cluster surprisal (CCS) for corresponding Sierpiński graphs in panel (a) as a function of $\beta$, which is the rate of error in memory when updating the mental model of the transition graph. The $\beta$ value at which the cross-cluster surprisal peaks is marked for all levels of the graph's hierarchy.}
	\label{Fig3.CCS_n}
\end{figure*}

\subsection{Ground Truth Graph Construction: Sierpiński Family}

Sequences of stimuli were drawn from walks on an underlying graph from the Sierpiński family. This section introduces the mathematical formalism for constructing Sierpiński family graphs.
The Sierpiński family is a graph generalization of the famous fractal fixed set Sierpiński triangle \cite{Survey}.
One major feature of the Sierpiński family is its recursive generation, in that each larger Sierpiński graph contains many smaller Sierpiński graphs, resulting in a self-similar pattern (see Fig. \ref{Fig3.CCS_n}(a) from left to right).
Because we aim to extend prior work that employed a modular graph \cite{ProbMotorSeqLearnability} featuring two hierarchical levels, here we opt for a family of graphs that has nearly the same degree for every node and a tunable number of hierarchical levels, on the backbone of a relatively simple and symmetric graph topology.
Notably, the self-similarity of the Sierpiński family can satisfy these aims.

We denote a generic unregularized Sierpiński graph with $S_{p}^{n}$,
where $p$ is the base, or number of nodes in the community at the finest level of the graph,
and $n$ is the power, or number of hierarchical levels.
We define our convention of level in a bottom-up manner.  The finest level is indexed by $1$ and the coarsest level is indexed by $n$.
The structure of the Sierpiński graph is self-similar: There are $p$ communities at each level of the graph, except that at level $1$ each individual node is a community. We expound the details on communities of the Sierpiński graph in the next section.
A mathematical definition of a generic Sierpiński graph is given as follows.
Each node in a Sierpiński graph $S_{p}^{n}$ has a unique index in the form of a natural number such that node $j$ belongs to an index set $\{0,1,2,...,p^{n}-1\}$.
We can also represent each index as a unique base-$p$ expansion written in the form $s_n...s_1|_p$
where $s_i\in \{0,1,2,...,p-1\}$ for any $i$.
The set of indices $V(S_{p}^{n})$ written as a base-$p$ expansion is then \cite{Survey}:
\begin{equation}\label{eq:SG_node}
    V(S_{p}^{n}) = \{ s_n...s_1 \mid{} \forall j \in \{1,...,n\}, s_j \in \{0,...,p-1\} \}.
\end{equation}
If there is an undirected edge between node $i$ and node $j$, we write $e_{ij|_{p}}\equiv(i,j)|_{p} \in E(S_{p}^{n})$, and the set of edges for $S_{p}^{n}$ are then defined using base-$p$ expansions of the node indices \cite{Survey}:
\begin{equation}\label{eq:SG_edge}
    \begin{split}
        E(S_{p}^{n}) = \{ & (sij^k,sji^k) \mid{} k \in \{0,...,n-1\}, \\
                          & s \in V(S_{p}^{n-k-1}), i,j \in \{0,...,p-1\} \},
    \end{split}
\end{equation}
where $sij$ and $sji$ are base-$p$ expansions of the part or the entirety of the decimal representations of the node indices, and $k$ indexes consecutive identical digits in a sequence which is abbreviated as $j^k$ (when $k=0$, $j^k$ is an empty sequence).

\subsection{Definition of Communities in a Sierpiński Graph}

A community---which in some contexts is interchangeable with ``cluster'' or ``module''---is often abstractly defined as a densely connected subgraph within a larger graph \cite{Communities}.
The graphs in the Sierpiński family have by definition a set of nested communities due to their self-similar construction. During graph construction, new hierarchical levels are added by making replicas of a seed Sierpiński graph created at the previous step.
This seed Sierpiński graph has its own community structure; hence, each replica will create a new community at each new hierarchical level. This construction process creates a nested community structure.
We will first define a notion of community in a Sierpiński graph and then explain how this definition fits the general definition \cite{Communities,community_detection} widely used in several contexts.

We define level-$l$, the most fine-grained community level, as a set of nodes $ s_n...s_{l+1}s_{l}...s_1 \in V(S_{p}^{n})$ that share the same leading $n-l$ digits in their base-$p$ expansions.
Thus, each level-$l$ community can be indexed by the truncated string $s_n...s_{l+1}$ (base-$p$ expansion); alternatively, one can use a decimal index instead, which is simply the base-$10$ expansion of $s_n...s_{l+1}$. 
As a result, there are $p^{n-l}$ communities at level-$l$ of the graph, such that each of the communities contains $p^{l}$ nodes.

Given a generic Sierpiński graph $S_{p}^{n}$, any hierarchical level $l$, and fixed parameters $p,n,l$, each of the $p^{n-l}$ communities has $p^1+...+p^l$ edges inside it, but only $p$ or $p-1$ edges connecting the nodes inside it to other communities.
Consequently, the number of within-community edges is in the order of $p^{l}$, which grows with level $l$, but the number of between-community edges is either $p$ or $p-1$, which stays constant.
Thus, the communities at any level of a generic Sierpiński graph $S_{p}^{n}$ are more densely connected inside each of them, compared to the sparse connections between them. This connection structure is consistent with definitions of community structure used in other contexts\cite{Communities,community_detection}.

\begin{figure*}[t]
    \centering
	\includegraphics[width=\textwidth, height=8.5cm]{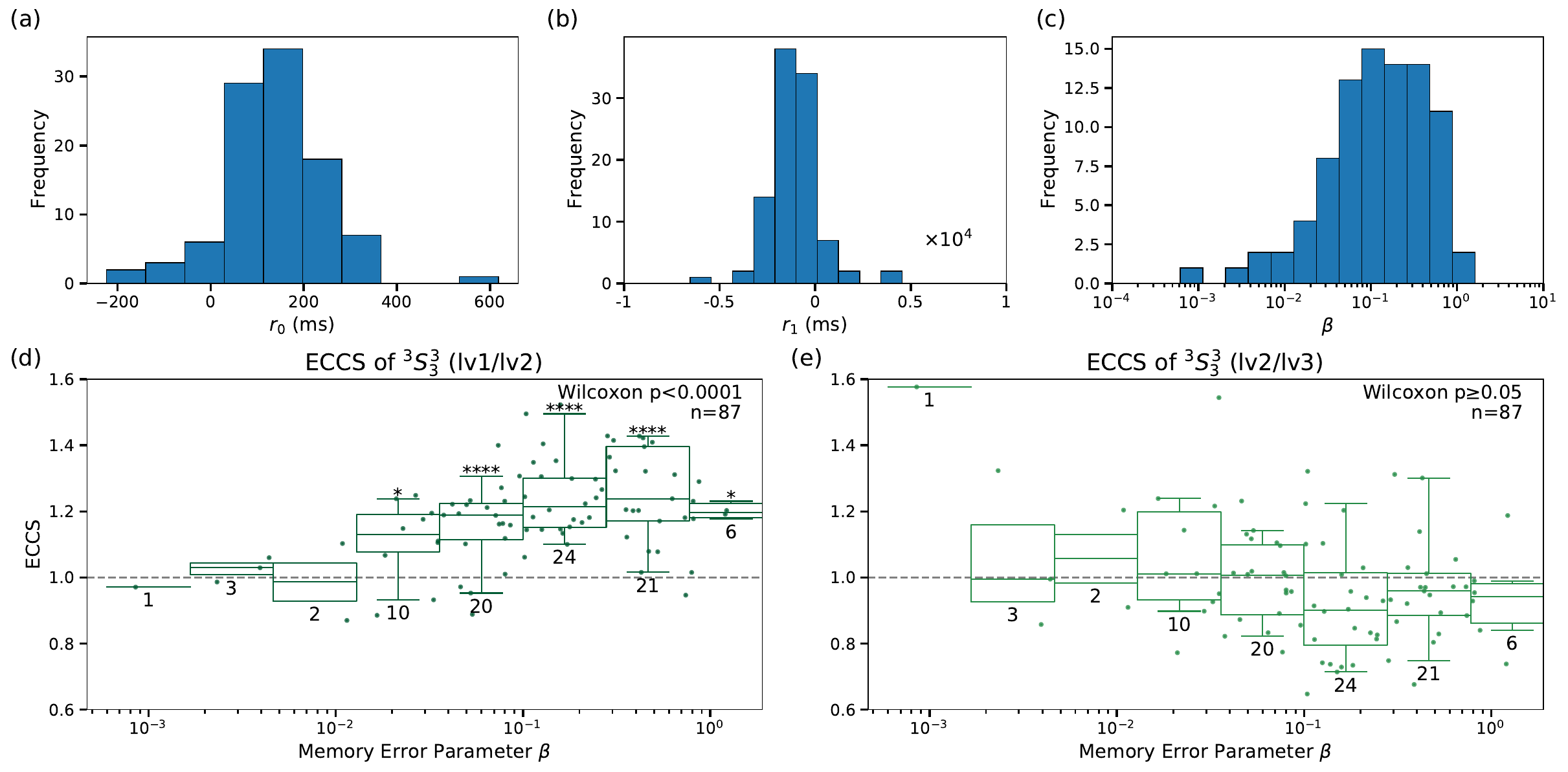}
        \caption{\textbf{Using the maximum entropy model to estimate the cross-cluster surprisal for individual human participants.}
        When fitting the maximum entropy model to the human reaction time ($rt$) data, we estimate three separate parameters as specified by the linear relation $rt = r_0 + r_1 a(\beta)$, where $r_0$ is intercept, $r_1$ is slope, and $\beta$ is the rate of error in memory when updating the mental model of the transition graph. 
        \emph{(a)} Histogram of the intercept $r_0$ in the linear model $rt = r_0 + r_1 a(\beta)$.
        \emph{(b)} Histogram of the slope $r_1$ in the linear model $rt = r_0 + r_1 a(\beta)$.
        \emph{(c)} Histogram of the $\beta$ values in the linear model $rt = r_0 + r_1 a(\beta)$. Note, here we only show data from participants whose $\beta$ satisfied $0 < \beta < \infty$; those excluded were $11$ participants whose $\beta=0$ and $2$ participants whose $\beta \rightarrow \infty$.
        \emph{(d-e)} Cross-cluster surprisal at level-$1$ \emph{(d)} and level-$2$ \emph{(e)}, including the $87$ participants whose $\beta$ satisfied $0 < \beta < \infty$.
        The $p$-values were obtained from one-sample Wilcoxon signed-rank tests, where we subtracted 1 from the cross-cluster surprisal value and compared the resultant number to a null distribution centered at zero.
        An individual asterisk above a boxplot indicates a $p$-value less than $0.05$; two asterisks indicate a $p$-value less than $0.01$; three asterisks indicate a $p$-value less than $0.001$; four asterisks indicate a $p$-value less than $0.0001$. Each one-sample Wilcoxon signed-rank test was performed on the data from a single $\beta$ bin. The $\beta$ bins are evenly spaced in logarithmic space and the definition of bins is the same throughout the analysis, except there are more bins in the simulations due to the $\beta$ range being larger in the simulations than in the experiment. The number below each boxplot is the number of human participants with $\beta$ values in that $\beta$ bin. For each bin, the box delineates the interquartile range whereas the bottom whisker delineates the $2.5\%$ percentile and the top whisker delineates the $97.5\%$ percentile.}
	\label{fig6.ECCS}
\end{figure*}

\subsection{Maximum Entropy Model}

The maximum entropy model, as used in prior work on modular graph learning \cite{MentalErrors,Lynn_info_process2020}, explains how humans may build their mental model of a transition graph as they react to transitions between two consecutive stimuli. This model captures systematic variation in reaction times through a time integration that balances the trade-off between \textit{expected recall distance} and \textit{recall inefficiency} (see \textit{SM} for details).
According to the model, learning occurs by estimating the transition probability for each edge on a transition graph $\mathbf{A}$ by normalizing mental counts of edge traversals (i.e., transitions):
$\hat{\mathbf{A}} = \Tilde{n}_{ij} / \sum_k {\Tilde{n}_{ik}}$
where $\Tilde{n}_{ij}$ is the mental count of transitions from node $i$ to node $j$.
Let $x_t$ be the index of the node visited at time $t$, then at time $t+1$,
the mental count is added to the edge that connects node $x_{t-\Delta t}$ and node $x_{t+1}$ where $\Delta t$ is drawn from a geometric distribution density function $f(\Delta t;\beta) = Ce^{-\beta \Delta t}$ parameterized by $\beta$:
\begin{equation}\label{eq:mental_counting}
\Tilde{n}_{x_{t-\Delta t},x_{t+1}}(t+1)=\Tilde{n}_{x_{t-\Delta t},x_{t+1}}(t) + 1,
\end{equation}
where $C$ is the normalizing constant that depends on whether it is in an infinite-time horizon or a finite-time horizon.
We call $f(\Delta t;\beta)$ the memory error distribution and its parameter $\beta$ memory error parameter because $f(\Delta t;\beta)$ affects the count $\Tilde{n}_{ij}(t+1)$ by placing non-zero weights on nodes visited many steps before.
Therefore, the last visited node $i=x_{t-\Delta t}$ could be any node visited in the random walk history, thereby rearranging the order of history when counting and creating ``errors.''
In the \textit{SM}, we explain why we elected to use the geometric distribution as the memory error distribution.

Given our model, we first define the underlying ground truth transitional graph $\mathbf{A}$ such that each entry $\mathbf{A}_{ij}$ corresponds to the probability that if node $i$ appears at time $t$, then node $j$ will appear at time $t+1$, for any non-negative integer-valued $t$.
We then consider two resulting learned transition probability matrices.
The first is a mean-field prediction in an \textit{infinite}-time horizon, and the second is a stochastic simulation in a \textit{finite}-time horizon.
The mean-field prediction in an infinite-time horizon takes the asymptotic form of $\hat{\mathbf{A}}=(1-e^{-\beta})\mathbf{A}(\mathbf{I}-e^{-\beta}\mathbf{A})^{-1}$.
To obtain the finite-time version of the learned transition probability matrix $\hat{\mathbf{A}}$, we instead use the finite geometric distribution $f(\Delta t;\beta)$ given permissible recall distances at each time-step, and normalize $\Tilde{n}_{ij}(T)$ after $T$ time-steps of learning in Eq. (\ref{eq:mental_counting}).
Refer to the original study \cite{MentalErrors} that devised the model for derivation details.

\subsection{Definition of Edge Levels and Cross-Cluster Surprisal in a Sierpiński Graph}

In this paper, we use the phrase ``cross-cluster surprisal'' (CCS) effect to describe a phenomenon that reaction times to transitions between-community are larger than reaction times to transitions within-community \cite{ProcessStructure}, and we use the term ``surprisal'' to indicate the slowing of reaction time reflective of people's expectations of a structure more broadly.
From a complementary perspective, the cross-cluster surprisal effect can also be defined based on the predicted mental representation of a two-level modular graph \cite{MentalErrors}.
Specifically, a cross-cluster surprisal effect occurs when the ratio of average within-community transition probabilities to average between-community transition probabilities is larger than one \cite{MentalErrors}.
Throughout the paper, we specifically refer to the ratio (not difference) of within-community to between-community transition probabilities as the ``CCS''; when the CCS is greater than 1, we say that there is a ``surprisal effect''.
In contrast to prior work, here we aimed to investigate surprisal effects on a hierarchical graph with more than two levels.
Thus, we generalized the notion of the CCS to any generic hierarchical graph with well-defined hierarchical levels.
To do so, first---similar to the community definition on nodes---we define level-$l$ ($l \in \{1,...,n\}$) edges as follows:
\begin{equation}\label{eq:SG_CCS_edge}
    \begin{split}
        E_l(S_{p}^{n}) \coloneqq \{ & (sij^{l-1},sji^{l-1}) \mid{} \\
                          & s \in V(S_{p}^{n-l-2}), i,j \in \{0,...,p-1\} \}.
    \end{split}
\end{equation}

Then a level-$l$ CCS (denoted as $\Delta_l$) where $l \in \{1,...,n-1\}$ takes the form:
\begin{equation}\label{eq:CCS}
    \Delta_l(\hat{\mathbf{A}}) \coloneqq \frac{ \frac{1}{|E_l|}{\sum_{(i,j)\in E_l}{\hat{\mathbf{A}}_{ij}}} }
    { \frac{1}{|E_{l+1}|}{\sum_{(i,j)\in E_{l+1}}{\hat{\mathbf{A}}_{ij}}} },
\end{equation}
where $\hat{\mathbf{A}}$ is the learned transition probability matrix.
Eq. (\ref{eq:CCS}) defines the CCS at level-$l$, which is the finer level of the two involved in the ratio.
In other words, for any edge level $l<n$, there is a corresponding CCS that compares level-$l$ edges to edges of the coarser-level $(l+1)$.

In all calculations of the CCS, we used the mental representation $\hat{\mathbf{A}}$
instead of the ground truth transition probability matrix $\mathbf{A}$,
because the CCS is a measure of the expected outcome and not of the ground-truth.

\subsection{Empirical Cross-Cluster Surprisal}

Here, to define the CCS from the empirical reaction time data (ECCS), we capitalize on the fact that the maximum entropy model estimates mental representations during its parameter fitting process.
The working definition of ECCS can be divided into three parts:
1) estimation of maximum entropy model parameters from reaction time data using gradient descent;
2) output of the last estimated mental count for each unique edge in the final iteration of the estimation process as carried out in step (1);
3) calculation of the CCS in the same way as done for simulation data where the simulated mental counts were used instead.
This working definition thus matches the manner in which we calculate the CCS in the simulation data as closely as possible, thereby making the comparison between simulation data and empirical data most meaningful.

Of note, all experimental analyses involving empirical cross-cluster surprisal were restricted to samples whose $\beta$ satisfies $0<\beta<\infty$. This experimental decision was taken in light of the intrinsic difficulties in differentiating $\beta=0$ from $\beta \rightarrow \infty$ by the reaction times of our participants alone. In both cases, the mental transition probability would in fact be the same for any experimental transition, thus resulting in the same reaction time. 

In addition, we observed that $11$ out of $100$ participants had a fitted $\beta=0$, which implies that they may have completed the task with maximum memory error. Further, $2$ out of $100$ participants had a fitted $\beta \rightarrow \infty$, which implies perfect memory.
Since these two numbers ($11$ and $2$) are much larger than the two tails of $0<\beta<\infty$ range in Fig. \ref{fig6.ECCS}(c), we excluded them from our analyses that involved $\beta$.

\subsection{Regularized Sierpiński Graphs}

It is known that humans are sensitive to local statistics; for example, humans react on average slower to nodes of higher degree than to nodes of lower degree \cite{ProbMotorSeqLearnability}.
In this study, we were specifically interested in how humans learn hierarchical structures. Here we deem the degree to be a confounding variable and thus we choose to modify Sierpiński graphs such that they become regular.
One way to regularize Sierpiński graphs is to add self-loops to the three boundary nodes (top, bottom left, and bottom right nodes in Fig. \ref{Fig2.CCS_p}(a) Left panel) of the unregularized graph. We denote this regularized graph as ${}^{3}S_3^3$, where the left superscript is the index of the regularization type.
In the \textit{SM}, we detail a list of regularization methods considered and the rationale of electing the self-loop approach.
We used ${}^{3}S_3^3$ as the ground truth graph for participants to learn in the probabilistic sequential motor task.
The transition matrix that prescribed the walk sequence on the regularized graph ${}^{3}S_3^3$ is the probability transition matrix whose entry is $1/3$ if there is an edge in ${}^{3}S_3^3$ and $0$ otherwise.
Since we rely on the definition of the edge level in order to calculate the CCS, we define the edges introduced in the self-loop regularization to have an undefined level, or level-$0$.

\subsection{Linear Mixed Effects Model}
As in our prior work \cite{MentalErrors}, we first filtered raw reaction time data to exclude the first $500$ trials, any trials during which participants' first attempts were incorrect,
and any trials during which reaction times were too short ($\leq 100 \textnormal{ ms}$) or too long ($\geq 3500 \textnormal{ ms}$) to capture reasonable motor reactions\cite{MentalErrors}.
We then fitted the filtered reaction times as a function of the transition type as well as covariates such as stimulus recency, time-steps, and keyboard input combinations, within a linear mixed effects model, whose formula in the standard R notation \cite{lme4} reads as follows:
\begin{equation}\label{eq:LME}
\begin{split}
    \textnormal{RT}\sim & \textnormal{ log(Trial)}+\textnormal{Target}+\textnormal{Recency}+\textnormal{Edgelv}+ \\
         & (1+\textnormal{log(Trial)}+\textnormal{Recency}+\textnormal{Edgelv}|\textnormal{ID},
\end{split}
\end{equation}
where ``RT'' is the reaction time, ``log(Trial)'' is the natural logarithm of trial number, ``Target'' is the keypress combination,
``Recency'' is the number of trials since the last occurrence of the stimulus during the current trial, ``Edgelv'' is the type of transition, and ``ID'' is the unique identifier for each of the $100$ participants in the experiment.
Because we were interested in comparing the reaction times for two adjacent edge levels, we used a custom dummy coding theme to convert the categorical variable ``Edgelv'' in Eq. (\ref{eq:LME}) into three binary variables ``lv01'', ``lv12'', and ``lv23'' (\textit{SM}, Table S1).

We fitted the Eq. (\ref{eq:LME}) model using the aforementioned dummy coding to the filtered reaction time data and computed the following statistics:
(i) reaction time difference between level-$1$ and level-$2$ transitions (``lv12'' in \textit{SM}, Table S1);
(ii) reaction time difference between level-$2$ and level-$3$ transitions (``lv23'' in \textit{SM}, Table S1).
Thus, coefficients for the binary variables ``lv12'' and ``lv23'' are average reaction time differences at level-$1$ and at level-$2$, after accounting for all other confounders and fixed effects.

\section{Results}

\subsection{Mean-Field Predictions Across Graph Bases and Graph Powers}
We employed the maximum entropy model \cite{MentalErrors} to predict a human's mental representations of Sierpiński graphs as a function of the memory error parameter $\beta$ (see Methods).
We considered Sierpiński graphs each with a base of three, four, and five, all having three hierarchical levels (Fig. \ref{Fig2.CCS_p}(a)).
In the infinite time limit, we find that the cross-cluster surprisal, that is how much the ratio of average mental transition probabilities at one level to those at the next coarser level is larger than one, displays a similar dependence on $\beta$ across all three graphs (Fig. \ref{Fig2.CCS_p}(b)).
The curve is unimodal at each level of the hierarchy, with no cross-cluster surprisal in the high- or low-$\beta$ limits.
Across all three graphs, the cross-cluster surprisal is stronger at the finer scale than at the coarser scale (Fig. \ref{Fig2.CCS_p}(b)).
As the base increases, the magnitude of the cross-cluster surprisal also increases at both fine and coarse levels of the hierarchy.
This behavior implies that transitions between large communities are more surprising than transitions between small communities.

To assess the generalizability of our findings, we next considered Sierpiński graphs with a power of three, four, and five, all sharing the same base of three (Fig. \ref{Fig3.CCS_n}(a)).
We again found that the cross-cluster surprisal displays a similar dependence on $\beta$ across all three graphs (Fig. \ref{Fig3.CCS_n}(b)).
As the power of the graph increases, we find that the strength of the cross-cluster surprisal tends to decrease, with the strongest effect at the finest scale (Fig. \ref{Fig3.CCS_n}(b)). We also observed a non-zero cross-cluster surprisal at each level of the hierarchy, suggesting that all levels of the Sierpiński graph ${}^{3}S_3^n$ can be learned given unlimited time.
When comparing hierarchical levels within graphs with a power of five, we found that the peak magnitude of the cross-cluster surprisal decreases first but later increases as the hierarchical level increases, and occurs at smaller values of $\beta$.
This behavior implies that in a community which is nested hierarchically with more than three levels, there may be a medium hierarchical level at which the transitions are least surprising when compared among the maxima across all levels.

\subsection{Stochastic Simulations of Different Sample Sizes and Walk Lengths}

Our results thus far are based on calculations that assume human learners have infinite time to learn.
To determine how our conclusions might depend on this assumption and the noise introduced with a finite sample size, we therefore now turn to simulations that use finite time and a finite sample size.
Our goal is to predict the likely outcomes of a laboratory experiment in which real humans spend finite time learning.
With that goal in mind, we recorded the mental counts of transitions on ${}^{3}S_3^3$ for simulated human learners across a range of $10$ possible $\beta$ values, each of which corresponds to the center of a log-uniformly spaced bin, and every bin has the same number of simulated human learners.
To implement each $\beta$ value, at each step of the random walk on ${}^{3}S_3^3$, the memory error size was drawn from a finite geometric distribution parameterized by $\beta$ \cite{MentalErrors} (also see Methods).
We then calculated the cross-cluster surprisal from the simulated mental counts and used statistical analyses to determine its significance.

\begin{figure*}[t]
    \centering
	\includegraphics[width=\textwidth, height=8.5cm]{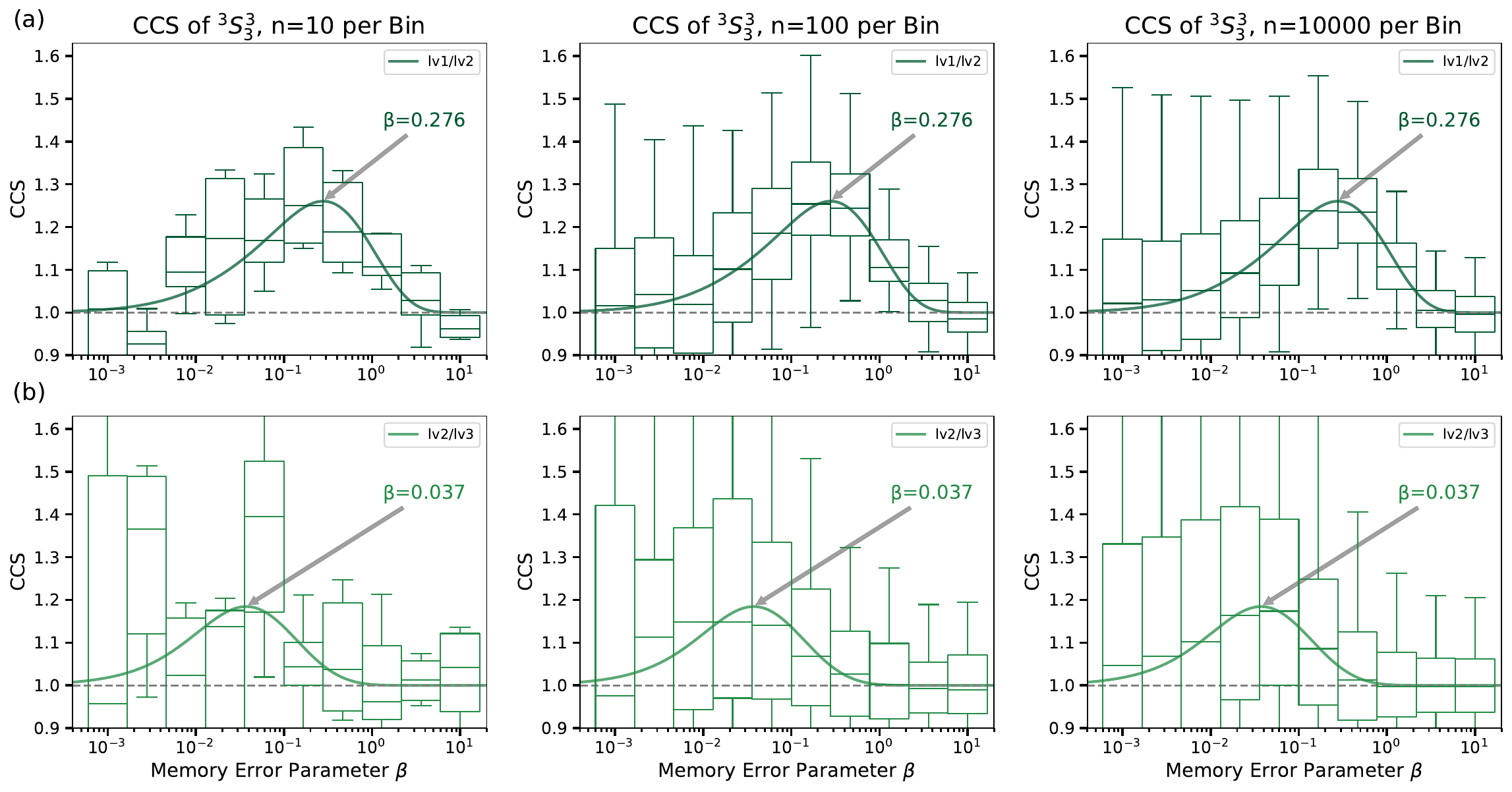}
	\caption{\textbf{Dependence of learning estimates on the number of simulated humans in the participant sample.} 
        Here we show boxplots of the cross-cluster surprisal for the Sierpiński graph ${}^{3}S_3^3$, across ten $\beta$ values and a walk length of $1500$ steps.
        For each $\beta$ value, the box delineates the interquartile range, the bottom whisker indicates the $2.5\%$ percentile, and the top whisker indicates the $97.5\%$ percentile. The solid curves are mean-field predictions of the cross-cluster surprisal at an infinite time horizon.
        We sampled $10$, $100$, and $10000$ agents per bin from the total of ten thousand available; columns differ by sample size.
	\emph{(a)} The cross-cluster surprisal (CCS) at the finer level of the hierarchy as a function of $\beta$: the rate of error in memory when updating the mental model of the transition graph. 
	\emph{(b)} The cross-cluster surprisal (CCS) at the coarser level of the hierarchy as a function of $\beta$. 
	}
	\label{fig4.stochasticCCS}
\end{figure*}

We found that the sample size affected the smoothness of the approximated distributions of the cross-cluster surprisal.
As the sample size increases, the mean of the distribution of cross-cluster surprisal values increasingly approximates the mean-field predictions (Fig. \ref{fig4.stochasticCCS}).
Intuitively, we also observed that the means of the distributions of adjacent bins are increasingly similar to each other as the sample size increases, indicating a growing smoothness.
Using $10,000$ simulated learners per $\beta$ bin,
we observed that for a walk length of $1500$, the cross-cluster surprisal at the finer level of the hierarchy is significant: that is, more than 97.5\% of the observed values are greater than the baseline for $\beta$ in $[0.1,1]$.
The peak cross-cluster surprisal is observed at $\beta=0.276$.
Using the same simulation setup, we observed that the cross-cluster surprisal at the coarser level of the hierarchy is not statistically significant: at most 75\% of the observed values are greater than baseline across the full $\beta$ range. The peak cross-cluster surprisal is observed at $\beta=0.037$.
This pattern of findings implies that the cross-cluster surprisal can be reliably detected at the finer but not coarser level of the hierarchy when the sample size is limited.

\begin{figure*}[t]
    \centering
	\includegraphics[width=\textwidth, height=8.5cm]{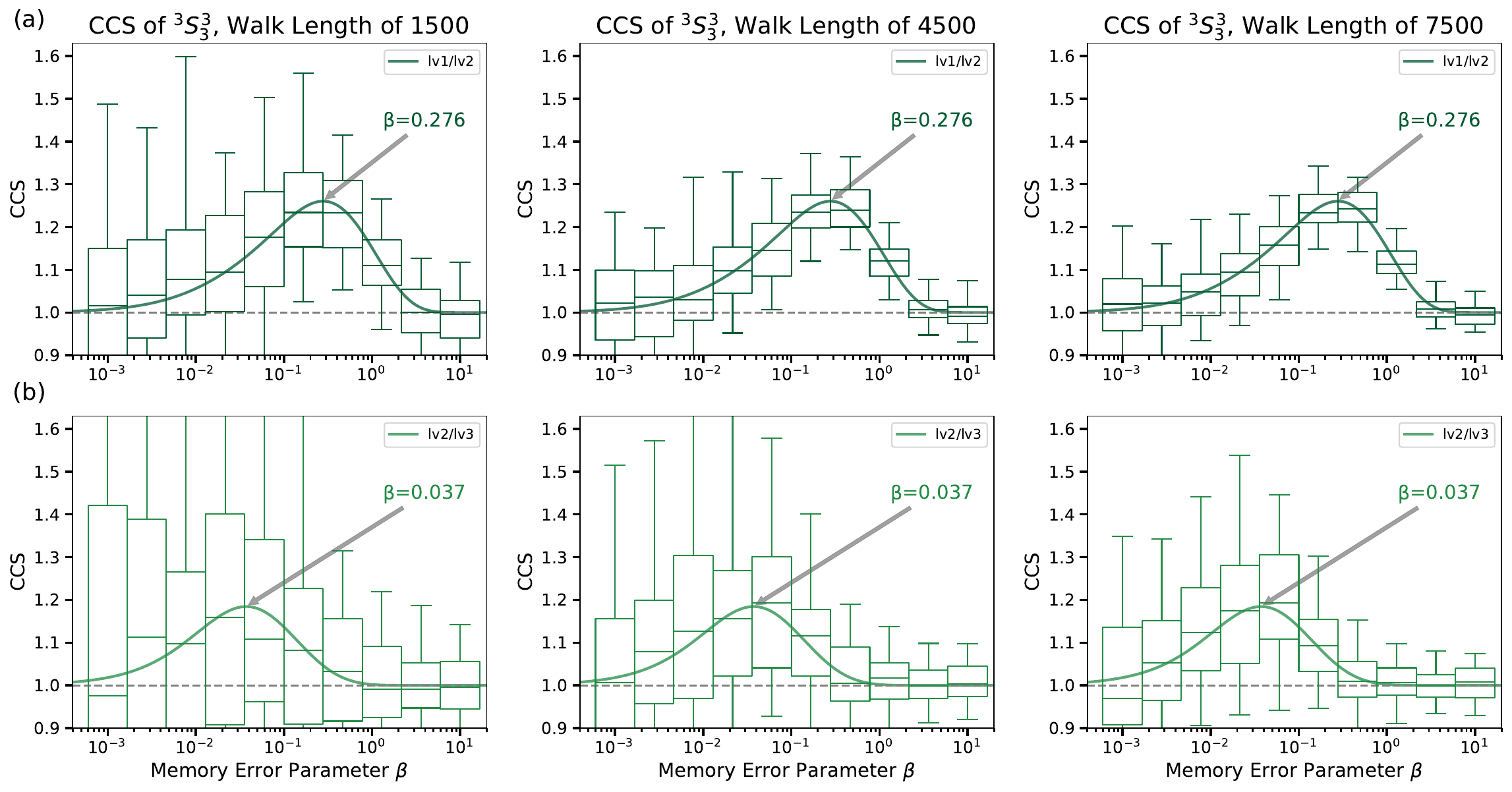}
	\caption{\textbf{Dependence of learning estimates on the number of simulated humans in the participant sample.} 
        Here we show boxplots of the cross-cluster surprisal for the Sierpiński graph ${}^{3}S_3^3$, across ten $\beta$ values and a sample size of 100 simulated participants. For each $\beta$ value, the box delineates the interquartile range, the bottom whisker indicates the $2.5\%$ percentile, and the top whisker indicates the $97.5\%$ percentile. The solid curves are mean-field predictions of the cross-cluster surprisal at an infinite time horizon. We sampled walk lengths of $1500$, $4500$, and $7500$ steps; columns differ by walk length.
	\emph{(a)} The cross-cluster surprisal (CCS) at the finer level of the hierarchy as a function of $\beta$: the rate of error in memory when updating the mental model of the transition graph.
	\emph{(b)} The cross-cluster surprisal (CCS) at the coarser level of the hierarchy as a function of $\beta$.
}
	\label{fig5.stochasticCCS}
\end{figure*}

Considering the time allotted for learning, we found that walk length affects the spread of the approximated distributions of the cross-cluster surprisal.
Using $100$ simulated learners per $\beta$ bin, we observed that the cross-cluster surprisal can be reliably detected at the finer level of the hierarchy for walk lengths as short as $1500$ steps (Fig. \ref{fig5.stochasticCCS}(a)).
However, the cross-cluster surprisal at the coarser level of the hierarchy could not be reliably detected, even for walks with $7500$ steps. Despite this prolonged exposure, more than 2.5\% of the observed cross-cluster surprisal values lay below the baseline; at a shorter walk length of $1500$ steps, more than 25\% of the the observed cross-cluster surprisal values lay below the baseline (Fig. \ref{fig5.stochasticCCS}(b)).
This observation suggests that a learning time of $1500$, which is on the scale of about half an hour or so, is sufficient for detecting the cross-cluster surprisal at the finer level reliably; but it may be insufficient to detect the coarser level surprisal even if the learning time is increased fivefold.

\subsection{Estimating the Surprisal Effect from Human Experiments}

Following our simulations, we next turned to laboratory experiments with real human participants. Our goal was to examine the presence and magnitude of the surprisal effect at two hierarchical levels in the Sierpiński graph ${}^{3}S_3^3$.
Accordingly, in an online laboratory platform, we presented $100$ human participants each with a sequence of $1500$ stimuli on their computer screen. The participants were asked to respond to each stimulus by pressing the corresponding keys (Fig. \ref{Fig0.schematic}). We recorded the reaction time for each stimulus to infer the participants' expectations about the transition probabilities among stimuli; a faster reaction corresponds to a more anticipated transition whereas a slower reaction corresponds to a less anticipated transition.

Previous empirical work in humans has shown that reaction times to within-community transitions are faster than reactions to between-community transitions \cite{ProcessStructure}, and the difference of the two is an empirical measure of the cross-cluster surprisal.
Notably, this difference in reaction times exists even after accounting for a set of covariates that may affect reaction times, such as the number of times the stimulus was observed in the last $10$ steps, the number of time-steps since the stimulus was last observed, and keyboard input combination differences that can drive biomechanical response differences \cite{ProcessStructure,ProbMotorSeqLearnability}.
In our experiment, we designed the graph such that it had four types of transitions: level-$i$ community transitions for levels $i=1,2,3$ and self-loop transitions. In line with previous literature \cite{ProcessStructure}, we included self-loop transitions to ensure that the graph was regular, such that each node had the same number of edges. See the \textit{SM} for additional details regarding our regularization procedure.

With this experimental design, we tested whether within-cluster transitions were statistically faster than between-cluster transitions. Specifically, we used a linear mixed effects model that accounted for the aforementioned covariates and a categorical variable (\textit{edge type}) that encoded the three hierarchical levels $i=1,2,3$ of the graph (see Methods).
We found that people tend to react faster to level-$1$ transitions compared to level-$2$ transitions by $21 \textnormal{ms}$ ($p<0.001$, $t$-test, \%$95$ CI: $[11, 30]$, DoF$=70598$).
Interestingly, reaction times to level-$2$ transitions were not statistically different from those to level-$3$ transitions
($p\geq0.05$, $t$-test, \%$95$ CI: $[-22, 21]$, DoF$=70598$).
In addition, we found that people tend to react faster to level-$1$ transitions compared to level-$3$ transitions by $20 \textnormal{ms}$ ($p<0.05$, $t$-test, \%$95$ CI: $[0, 41]$, DoF$=70598$).
These data indicate the existence of a statistically significant surprisal effect at the finer scale of the graph but not at the coarser scale.

In addition to measuring the surprisal effect from reaction times at a group level, we investigated the surprisal effect from estimated mental representation at an individual level. Specifically, using the maximum entropy model \cite{MentalErrors}, we estimated a mental representation for each individual participant (Fig. \ref{fig6.ECCS}(a-c)).
We then calculated cross-cluster surprisal values based on the estimated mental representations (ECCS; see Methods).
By this measurement, we found a significant cross-cluster surprisal at level-$1$ ($p<0.001$, $W=3717$, one-sided Wilcoxon signed-rank test, $n=87$).
Next we asked whether the cross-cluster surprisal was greater for people with higher versus lower values of $\beta$. After separating the data into discrete $\beta$ bins, we found that the level-$1$ cross-cluster surprisal was significant for intermediate $\beta$ values ($\beta\in$ $[0.1,1]$; Fig. \ref{fig6.ECCS}(d)).
Interestingly, these effects were not observed at the coarser scale of hierarchy in the graph. We did not observe a significant cross-cluster surprisal at level-$2$ ($p\geq0.05$, $W=1483$, one-sided Wilcoxon signed-rank test, $n=87$). After separating the data into discrete $\beta$ bins, we found that the level-$2$ cross-cluster surprisal was not significant for any $\beta$ values (Fig. \ref{fig6.ECCS}(e)).
Taken together, both group-level results from the linear mixed model and individual-level results from the maximum entropy model indicate that the surprisal effect is easily detectable at level-$1$ but not at level-$2$.

\subsection{Factors Impacting the Surprisal Effect Estimation in Human Experiments}

\begin{figure*}[t]
    \centering
	\includegraphics[width=\textwidth, height=8.5cm]{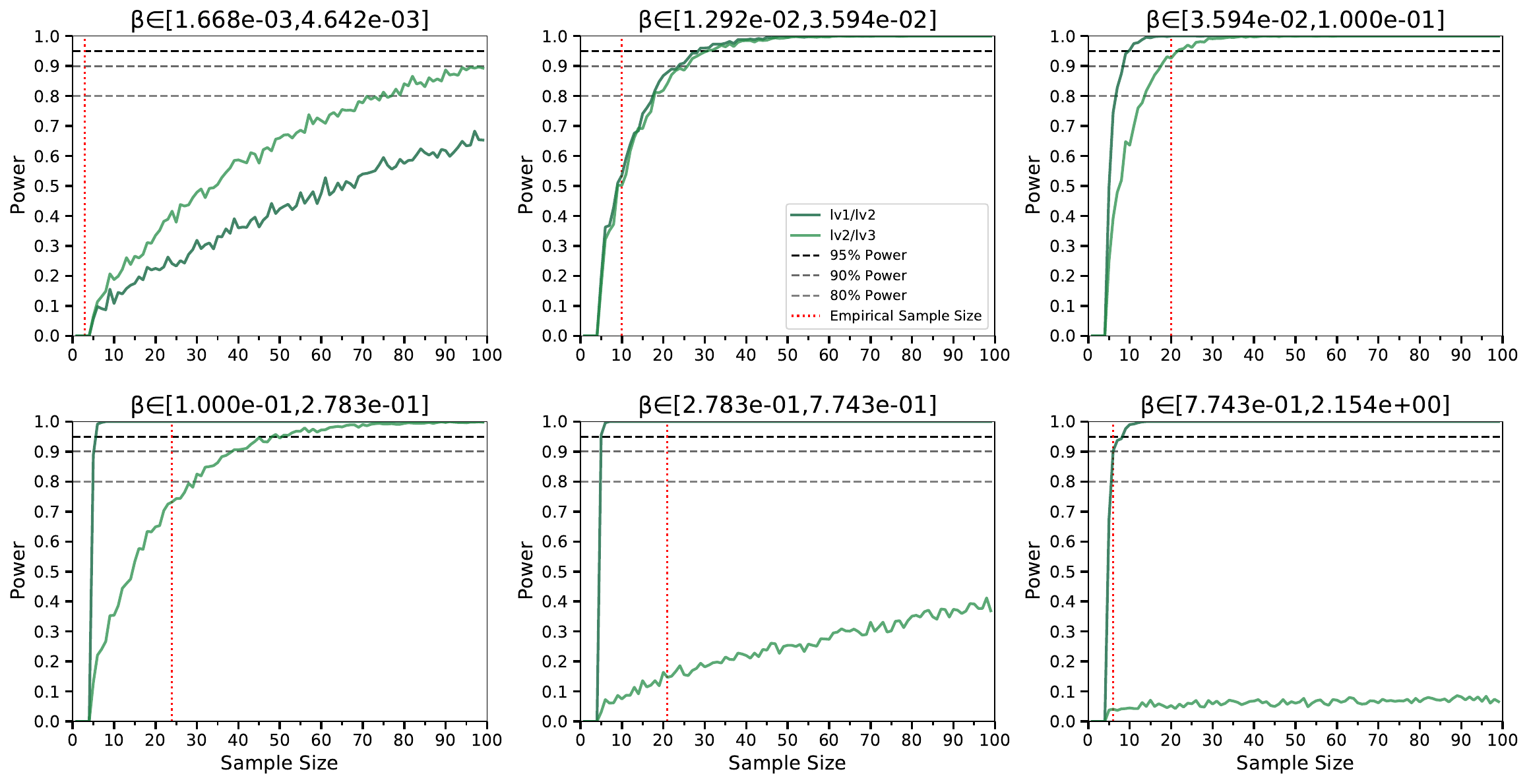}
        \caption{\textbf{A power analysis for estimating the surprisal effect in human experiments.}
        Here we provide plots of powers of one-sided Wilcoxon signed-rank tests on simulated data obtained from the same Sierpiński graph used in the experiment (${}^{3}S_3^3$).
        To the leftmost $\beta$ bin in Fig. \ref{fig4.stochasticCCS}, Fig. \ref{fig5.stochasticCCS}, and Fig. \ref{fig6.ECCS} we assign an index of one, and to the second leftmost bin we assign an index of two, and so on and so forth. Hence, the $\beta$ bin indices in the plots here refer to the corresponding $\beta$ bins in Fig. \ref{fig4.stochasticCCS}, Fig. \ref{fig5.stochasticCCS}, and Fig. \ref{fig6.ECCS}.
        Note that we only included $\beta$ bins whose empirical sample size (as shown in Fig. \ref{fig6.ECCS}) is greater than two. To estimate the power of the one-sided Wilcoxon signed-rank test given a sample size $n=1,2,...,99$ for each $\beta$ bin, we uniformly sampled $n$ agents with replacement from the simulation data that had a total of $10,000$ agents per beta bin. We then repeated this process $1000$ times.
        Next, we approximated the statistical power by calculating the ratio of repetitions in which the one-sided Wilcoxon signed-rank test yielded a $p$-value that was less than $0.05$.
        Because the surprisal effect can happen at two hierarchical levels in the Sierpiński graph ${}^{3}S_3^3$, here we show power estimates for both levels (shades of green), with different power baselines (95\%, 90\%, 80\%; dashed lines) for reference.
	}
	\label{fig7.powers}
\end{figure*}

Our mean-field results at the infinite time limit indicated that the cross-cluster surprisal existed at both finer and coarser hierarchical scales of the graph.
However, our finite-time simulations indicated that the surprisal was easily detectable at the finer scale and less detectable at the coarser scale, and depended on both the walk length and sample size.
In our human experiments, which spanned finite time and employed a small sample, we found that the cross-cluster surprisal was significant at the finer scale of the Sierpiński graph ${}^{3}S_3^3$ but not at the coarser scale.
We hypothesized that the non-significant surprisal at the coarser scale was in part due to a lack of statistical power arising from the finite sample size in the human experiments.
To test this hypothesis, we carried out a power analysis (Fig. \ref{fig7.powers}).
We found that the power to detect the coarser-level cross-cluster surprisal at the empirical sample size for each $\beta$ bin was below $80\%$ for all but the fifth $\beta$ bin.
Even for the fifth bin, the power was below $95\%$, meaning there was still a $\>5\%$ chance that we would not detect a significant cross-cluster surprisal effect.

\begin{figure}[t]
    \centering
	\includegraphics[width=8.6cm]{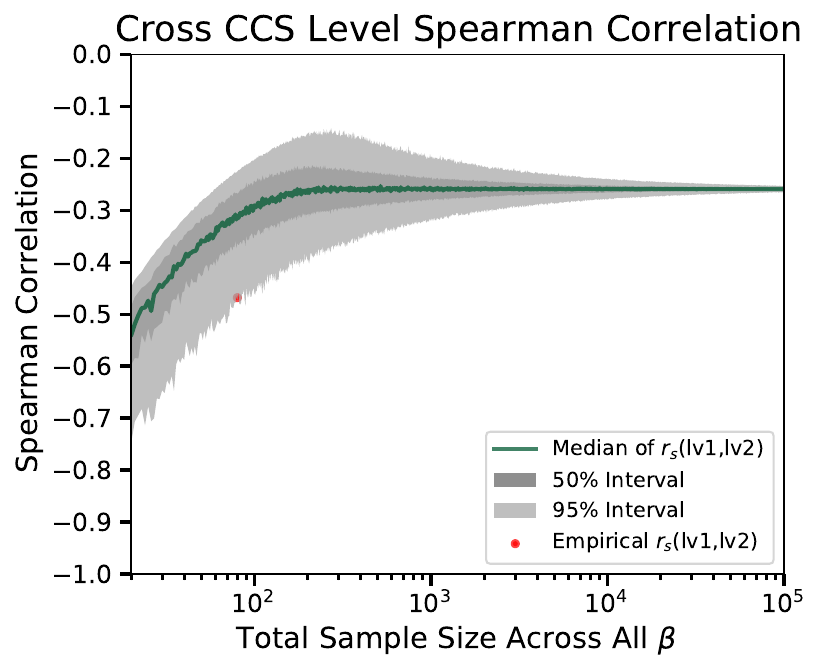}
	\caption{\textbf{Trade-off in learning finer versus coarser scales of hierarchical graphs.}
        Here we plot the Spearman correlation coefficient between the cross-cluster surprisal at the finer scale and the cross-cluster surprisal at the coarser scale, across all $\beta$ bins for both simulations (in grey) and the experiment (in red).
        To estimate the spread of Spearman correlation coefficients at different values of $N$ (the total sample size) for the numerical simulations, we uniformly sampled $N$ agents with replacement from the simulation data which had a total of $10,000$ agents per beta bin. Then, we calculated the Spearman correlation coefficient between the cross-cluster surprisal at the finer scale and the cross-cluster surprisal at the coarser scale.
        Finally, we repeated this process $1000$ times. From the $1000$ resultant estimates of the Spearman correlation coefficients for each $N$, we calculated the median, $50\%$ interval, and $95\%$ interval. The red dot indicates the Spearman correlation coefficient ($r_s=-0.468$) for the empirical data in Fig. \ref{fig6.ECCS}.
	}
	\label{fig8.corr}
\end{figure}

In addition to performing a power analysis, we examined whether the humans who learned the finer level also tended to not learn the coarser level.
If this was the case, then it would suggest the existence of a trade-off in learning, such that humans may devote more mental resources to learning one level of the hierarchy to the detriment of other levels.
To examine this possibility, we calculated the Spearman correlation coefficient between the cross-cluster surprisal detected at the finer level of the hierarchy and that detected at the coarser level of the hierarchy.
To determine whether the measured correlation was greater than expected in non-human agents, we estimated the cross-cluster surprisal from numerical simulations with $10,000$ agents per bin and a walk length of 1500.
As shown in Fig. \ref{fig8.corr}, for a given sample size, the Spearman correlation coefficient varied greatly in non-human agents, and this variation only diminished appreciably at a sample size of about $10^4$. At the empirical sample size of $87$, about $95\%$ of the simulated Spearman correlation magnitudes were smaller than that observed in the humans (Fig. \ref{fig8.corr}).
As predicted by the model, learners with a stronger cross-cluster effect at finer scales have a weaker cross-cluster effect at coarser scales, and \emph{vice versa}.
However, as shown in Fig. \ref{fig8.corr}, this trade-off in learning is greater in humans than in roughly $95\%$ of the simulated agents, suggesting that humans may devote mental resources to learning one scale of the graph more than another, rather than distribute those mental resources equally among all scales of the graph.

\section{Discussion}

Statistical learning of transition structures manifests in multiple aspects of human life, from learning visual patterns \cite{SLVisual} to learning a language \cite{AcqLang}.
Prior statistical learning studies have demonstrated that humans can learn topological features of transition graphs.
Some transitions graphs are more learnable than others \cite{qian2022optimizing}, and even a single graph can be learned differently by different individuals, as evidenced by variations in their mental representations.
The specific topological features that humans can learn include degree \cite{SLInfant,ProbMotorSeqLearnability}, community structure \cite{ProcessStructure,ProbMotorSeqLearnability,MentalErrors}, and betweenness centrality \cite{ProbMotorSeqLearnability}, hence spanning from local to meso-scale to global structure.
Notably, humans are also sensitive to transitions that do not fit the statistics of the learned structure, and such sensitivity depends on precisely how the learned statistics have been violated \cite{MentalErrors}.

Prior studies of graph learning have typically examined graphs with community structure that exists at a single scale \cite{ProcessStructure,ProbMotorSeqLearnability,MentalErrors}.
Graphs containing hierarchical communities have not yet been examined.
Accordingly, here we innovate by investigating how humans learn graphs with hierarchical community structure.
Specifically, we employed regularized Sierpiński graphs, which are (i) symmetric, thus relatively easy to analyze, and (ii) small, thus suitable to be employed in sequential motor learning tasks designed for humans.
Based on the self-similar topology, we can have well-defined hierarchical levels on edges as well as nested communities for Sierpiński graphs, which can be regularized such that every node has the same number of edges, consistent with prior studies \cite{ProcessStructure,ProbMotorSeqLearnability,MentalErrors}.

Using a finite time horizon, our simulations indicated that the cross-cluster surprisal is consistently detectable at the finer level of the hierarchy, but less detectable at the coarser level of the hierarchy. 
Turning from simulation experiments to human experiments, we collected reaction time data from humans as they learned a specific type of Sierpiński graph (${}^{3}S_3^3$) on an online platform.
In these experimental data, we similarly observe that the cross-cluster surprisal is detectable at the finer level of the hierarchy, but not at its coarser level.
Interestingly, we observe a strong negative correlation between the cross-cluster surprisal at the finer versus coarser levels of the hierarchy, indicating that human participants who learned the finer level well tended to learn the coarser level less well and vice versa.
This trade-off in learning has not previously been observed and is likely underestimated by our maximum entropy model.
Taken together, our data reveal three factors that could decrease the capacity to observe a significant cross-cluster surprisal at the coarser scale of the hierarchy:
limited statistical power due to the number of participants, limited learning time, and a trade-off between the learning of one hierarchical level and the learning of another hierarchical level.

\subsection{Learning Sierpiński Graphs with Infinite Time}

Our study began by considering simulations of human learning using a well-validated maximum entropy model \cite{MentalErrors} of human behavior. Our goal was to provide mean-field solutions to the learning problem in the infinite time limit.
As a measure of learning, we used the so-called cross-cluster surprisal, which measures the slowing of reaction times at boundaries between clusters in the graph. We found that the cross-cluster surprisal increased as the base for the Sierpiński graph increased.
Specifically, we observed that, during a random walk on a base-$p$ graph, a walker in a given cluster at node $i$ had a $(p-1)/p$ chance of staying inside the cluster for almost all $i$. The effect of this relation is to increase the average number of consecutive steps taken in a given cluster before transitioning to a new cluster, as $p$ increases. Because the human errors in memory of these transitions smoothly decrease with time in the past, most errors will tend to swap two nodes seen near-in-time, which for high $p$ will also tend to be two nodes in the same cluster. The cumulative effect of this process is that humans will tend to over-estimate the probability of staying inside a cluster and under-estimate the probability of moving to a new cluster, hence increasing the magnitude of the cross-cluster surprisal effect.

There are multiple scales of the cross-cluster surprisal effect on a hierarchical graph, and the memory error parameter $\beta$ affects these scales.
Specifically, as the memory error parameter $\beta$ decreases, \textit{more} errors will tend to swap two nodes seen far-in-time (as opposed to near-in-time), shifting the cross-cluster surprisal to coarser scales.
Holding the base $p$ constant, this shift to a coarser scale will result in a sensitivity to higher hierarchical levels, which is evidenced in the shifting peaks in the cross-cluster surprisal curves on power-$n$ graphs.
As such, learners with a small $\beta$ should tend to learn the transitions at a higher hierarchical level better than those at a lower hierarchical level, resulting in a more pronounced cross-cluster surprisal effect at a coarser scale.

\subsection{Cross Cluster Surprisal in a Finite Time Horizon is Approximated through Simulations}

The quality of learning large graphs depends crucially on the learning time, and for small learning times shows significant stochastic variation due to the random walk realizations \cite{Klishin_exposure2022,Klishin_learn2022}.
As such, we next assessed how the amount of time allocated to the learning process would affect the detection of the cross cluster surprisal.
Our simulation analyses across a range of finite time horizons indicated that cross-cluster surprisal was consistently detectable at the finer scale of the hierarchy (level-$1$), but not at its coarser scale (level-$2$).
Such simulation-based findings, in turn, indicate that an experimental design consisting of a $1500$-step random walk and $100$ learners may not be sufficient to detect coarser-scale surprisal. Our laboratory experiment in human participants confirmed the latter observation.

Although our simulations corroborated our experimental findings, some aspects of our simulation analyses may benefit from further development.
First, we only assessed the influence of walk length on learning by simulating mental transition counts. In contrast, in our empirical data we inferred surprisal from reaction times in the context of sequential tasks, but have no direct access to the mental counts.
Future work could develop a more precise behavioral signature of human graph learning that may be concomitantly applied to both simulated and empirical scenarios.
Second, some factors that are unique to human experiments may affect learning and increase the variance in human reaction time data but not simulation data, such as variations in the baseline dexterity of a given participant and their prior experiences with sequential motor tasks.
Moreover, variability may exist in the mechanics of specific keyboard combinations, whereby a given participant may unpredictably find some to be easier to learn than others. Accounting for all the above mentioned factors in the context of a simulation is not currently feasible.
Thus, the noise estimates in the simulation framework are conservative estimates of the noise that may be found in the empirical data. These and related considerations may inform behavioral neuroscience work that combines simulation and experimental study paradigms.

\subsection{Detectability of the Surprisal Effect and Trade-Off Between Finer-scale and Coarser-scale Transitions}

Our experimental findings in human participants highlighted significant group- and individual-level cross cluster surprisal effects for finer-level transitions, thereby replicating evidence from previous work that employed modular graphs \cite{ProcessStructure,ProbMotorSeqLearnability,MentalErrors}.
By contrast, we did not observe significant cross-cluster surprisal for coarser-level transitions.
A potential determinant of such phenomenon could be the lack of statistical power of our empirical setup. We tested this hypothesis with follow-up analyses, which showed that, despite the large number of participants and ample learning time, our experiment in human participants was underpowered to detect coarser-level cross-cluster surprisal for virtually all $\beta$ bins.
We conclude that a larger participant sample and, possibly, longer learning time may increase the likelihood of detecting cross-cluster surprisal effects at the coarser hierarchical scale, and could be fruitfully implemented in future studies.

To further explore our coarser-level findings, we also characterized the relationship between cross-cluster surprisal at the finer level and that at the coarser level.
Correlation analyses indicated a statistically unlikely strong negative association between the two variables, whereby participants who learned finer-level transitions well performed worse than typical stimulated agents at the higher hierarchical level, and \emph{vice versa}.
This inverse relationship has relevant neurobiological implications, and suggests that a trade-off process during learning, likely owing to finite capabilities of the human mind, may represent a signature of learning of hierarchical topology in real-world, time-constrained scenarios.
Specifically, humans allocating a high level of mental resources to learn one hierarchical level of the graph may do so at the expense of learning at another hierarchical level.
This finding suggests a potential trade-off between robustness to noise---where noise differs by hierarchical level---and flexibility to learn multiple levels; we note that both robustness and flexibility are properties of a goal-directed system \cite{trade-offs2018}.
Such an imbalanced allocation of brain resources could be implemented by variations in attention, or driven by perceived differences in the value of fine versus coarse patterns of information \cite{attention2015}.
Future work may benefit from multimodal experimental designs, relying on combined behavioral and, likely, functional imaging measures, to capture the underlying neural processes.

\subsection{Drivers of the Surprisal Effect and Implications for Future Work}

In addition to the possibility of a learning trade-off between different hierarchical levels, the learnability of cross-cluster transitions at finer hierarchical scales may also be related to the topological properties of graph $^{3}S_{3}^{3}$ and to the random walk scheme of graph learning. Because the number of edges exponentially decreases as the level of the edge increases and because each edge is equally likely to be traversed on a random walk on such graph, one would only expect a $1/14$ chance that the next step traverses any one of the level-$3$ edges, similar to self-loops. This pattern, in turn, naturally leads to a limited number of visits on high-level transitions, as opposed to lower level transitions. Thus, we conclude that the intrinsic organizational properties of hierarchical graphs, such as those implemented here, may be an additional driver of our cross-cluster surprisal findings. As discussed above, manipulating experimental conditions by, for instance, introducing longer random walks in the context of finite-time experiments, may attenuate noise effects arising and would be beneficial in future work.

Our simulation and experimental findings confirmed that time limits negatively influence the likelihood of detecting a significant cross-cluster surprisal effect at coarser levels of the hierarchy.
The required learning time and sample size can be estimated with calculations from the recently introduced exposure theory of graph learning \cite{Klishin_exposure2022,Klishin_learn2022}. While exposure theory originally aimed to predict edge learning in finite time at a binary level, the formalism can be extended to estimate the contrast between the learned transition probabilities, and thus the cross-cluster surprisal at different hierarchical levels.
Consequently, future experimental work in humans could consider allocating more time for graph learning, in addition to including a larger participant sample size.
In so doing, longer walks would lead to an increase in the raw counts of coarser-level transitions, thereby enhancing the detectability of the second-level cross-cluster surprisal. For a more parsimonious evaluation of level-2 transitions, one option would be to consider a modification of the experimental paradigm to attain a systematic increase in the probabilities of coarser-level transitions. A second option would be to add a flag to the stimulus presentation to indicate to the participant that they are about to experience a coarse-level transition. This explicit flag could improve the participant's ability to differentiate between coarse- and finer-level transitions. Using either option, one could proactively facilitate the detection of coarser-level transitions, and could also empirically validate our hypothesis that learning coarser-level transitions, in the context of the current experimental design, is impaired owing to a saturation of human neural resources for finer-level transitions.

Finally, we previously discussed that an appropriate graph to investigate hierarchical learning ought to have at least three hierarchical levels, to allow for the existence of at least two cross-cluster surprisal scales. The graph $^{3}S_{3}^{3}$ used in our study satisfies the requirements of a relatively small size and self-symmetry. However, we note that $^{3}S_{3}^{3}$ is still considerably larger than the modular graph used in previous experiments \cite{ProcessStructure,ProbMotorSeqLearnability,MentalErrors}. This experimental feature, in turn, places greater cognitive demands on the participants, potentially leading to decreased task performance and more frequent attentional lapses. In addition, every edge in $^{3}S_{3}^{3}$ is traversed overall $28$ percent less frequently than in the previously used modular graph. On balance, both the above characteristics are likely to have detrimentally influenced cross-cluster surprisal effects at the coarser scale in our human experiment. Collectively, our study paves the way for future investigations of human learning in hierarchical graphs, and offers important pragmatic considerations in the context of the experimental paradigms that may be best suited to investigate these.

\subsection{Future Directions}
Our study represents the first attempt to understand human hierarchical graph learning and to test the hypothesis that human learners exhibit cross-cluster surprisal effects at more than two hierarchical levels in a graph during a sequential motor learning task. To advance research in this field, future work may benefit from varying the hierarchical structures of a graph to a greater degree. One possible strategy is to remove the symmetry requirement that was a cardinal component of our experimental design. Another area of improvement pertains to the graph size. Our Sierpiński graph had a limited number of nodes, whereas real-world networks such as Wikipedia networks \cite{Hier_wiki} and the Semantic Web \cite{Hier_MoreNets} that humans are exposed to are likely to possess a significantly larger number of nodes. Thus, further experimental designs could implement scalable learning tasks that incorporate significantly larger graphs with different types of hierarchical structures. Finally, further advancements in the formulation of our current mental model \cite{MentalErrors} of graph learning may help refine our hypothesis that  humans may exhibit a learning trade-off that could favor finer-level transitions at the expense of coarser-level transitions.

\subsection{Conclusion}
In conclusion, our study combines simulation-based data and an experimental graph-learning paradigm administered to human participants. Our findings establish that finer-level transitions on a hierarchical graph, measured with the cross-cluster surprisal metric, are more easily detectable than coarser-level transitions. We also observe a strong negative correlation between cross-cluster surprisals at fine versus coarse scales, suggesting the existence of a trade-off in human learning, whereby the learning accuracy for one class of transitions may be maximized at the expense of the other one.
For hierarchical graphs, learning time and sample size are potential additional determinants of the detectability of cross-cluster surprisal at coarser scales.

\begin{acknowledgments}
This research was funded by
the Army Research Office (DCIST-W911NF-17-2-0181)
and the National Institute of Health (R21-MH-106799).
The content is solely the responsibility of the authors and
does not necessarily represent the official views of any of
the funding agencies.
We thank David Lydon-Staley for providing help in formulating demographic questions,
and thank Mathieu Ouellet for helping to find the potential transition graph candidates.
The computational workflow and data management for simulations used in this work was supported by the signac data management framework \cite{ADRG18,RAD+18}.
The color maps for visualizations were supported by the ColorBrewer2 tool \url{https://colorbrewer2.org/}.
\end{acknowledgments}

\section*{Author Contribution}

X.X. carried out the experiment and simulations, and wrote the paper;
A.A.K. contributed to the main theoretical analysis of the work;
J.S. contributed to the main design of the experiment and main data analysis;
C.W.L. contributed to early conceptual development and the code used in the work;
A.E.K. contributed to the experiment;
L.C. provided extensive help in the revisions;
D.S.B. guided and supported the work, and edited the paper.

\nocite{*}
\bibliography{main10.5}

\end{document}


\title{Supplemental materials to the manuscript}
\maketitle

\section{Maximum Entropy Model}

We elected a geometric distribution as the scrambling distribution because the geometric distribution optimizes a trade-off between two terms known to be relevant for learning \cite{MentalErrors}:
\textit{expected recall distance} and \textit{recall inefficiency}. These two variables could also reasonable be referred to as the ``error of a recalled stimulus'' and the ``computational complexity'', respectively, as in Ref. \cite{MentalErrors}.
According to the scrambling assumption, the source node for the transition can be ($\Delta t + 1$)-steps away from the target node during mental counting, where $\Delta t=0,1,...$ corresponds to the recall distance.
The memory scrambling distribution assigns a probability to recall distance, and the expected recall distance $\mathbb{E}(\Delta t)$ is the expectation of recall distance in the scrambling distribution.
The recall inefficiency $\mathbb{E}(\textnormal{log}f(\Delta t;\beta))$ is conceptually similar to the definition of expected recall distance, but corresponds instead to the expectation of the logarithm of the probability for each recall distance, rather than to the recall distance itself.
Thus, if there is only one recall distance with all other choices having a probability equal to zero, then the recall will be demanding or inefficient, as one will have to select only the stimulus at that exact recall distance;
conversely, if all recall distances are equally likely, the recall can be any stimulus visited in the past, thus being easy or efficient.

Hence, the expected recall distance and recall inefficiency comprise a trade-off:
If the recall is very efficient, the expected recall distance will be large, which in turn lowers the accuracy of recalling the last visited stimulus that, by definition, has a recall distance of zero.
The total resource cost, then, corresponds to a weighted sum of expected recall distance and recall inefficiency: $\beta \mathbb{E}(\Delta t) + \mathbb{E}(\textnormal{log}f(\Delta t;\beta))$.
Here, we maintain that the brain minimizes this total resource cost, as argued in prior work \cite{MentalErrors}.
In the total resource cost trade-off, the factor $\beta$, which we call the ``scrambling parameter'' throughout the paper, corresponds to how much weight the brain puts on accuracy over efficiency \cite{Brain_Tradeoff,MentalErrors}.
The minimization of the total resource cost yields a geometric distribution for the scrambling distribution $f(\Delta t;\beta)$ as the solution.

When normalizing the mental counts to derive the mental probabilities for transitions between stimuli, there is a choice of what the mental counts are before the learners start the graph learning task. In this paper, we did not assume that the learners have any prior expectations of the graph size or possible transitions of the underlying graph they were to learn, and therefore the mental counts are zero at the beginning. In the case of division by zero during normalization, we assigned a value of zero to the mental transition probability.

\section{Experimental Design on Graph Choice}
Within the Sierpiński graph family,
there are several options regarding the ground truth graph that governs the transition probability of the stimuli in the experiment.
We first enumerate some guidelines on several empirical aspects of learning a graph,
which constrain the graph choice for the experiment.
By violating the guidelines on those empirical aspects, we shift the difficulties in experimental design to data analysis.
To simplify the analysis, we attempt to make the design satisfy as many guidelines as possible.
We then evaluate a number of graphs in the Sierpiński graph family to choose one of them that best meets the guidelines.

\subsection{Learning Time and Visit Frequency}
Since the nodes are sampled according to a random walk on the graph,
how long the experiment lasts is directly related to the count with which each node is visited.
For a random walk in infinite time, the probability of landing on any node or edge is proportional to its degree; thus, we assume for simplicity that the probability at its limit is a good estimator on the expected count of visiting a node or edge when the walk length is at least $1500$.
To ensure that the count is sufficient, we define sufficiency based on the estimated expected count of $100$ $\unit{visits} / \unit{node}$ or $50$ $\unit{visits} / \unit{edge}$
in the previous experiment where a $1500$-step random walk was performed
on a $15$-node modular graph in $30$ minutes \cite{MentalErrors}.
\textbf{Guideline:} Thus, about an hour or $3000$ visits on nodes on a regular graph
with the estimated expected count of at least $100$ $\unit{visits} / \unit{node}$ would allow the graph to have at most $30$ nodes.

\subsection{Task Complexity}
We used the same probabilistic sequential motor task used in our previous study
to keep the experiment setting as similar as possible and to avoid confounding variables.
The complexity is considered under the context of learning within an hour.
The task design in the previous study only involved one hand and one- or two-finger combinations,
resulting in $15$ combinations mapped uniquely to $15$ nodes in the modular graph \cite{MentalErrors}.
To study hierarchical effects, we need to have at least three levels, whereas the modular graph has only two levels.
The simplest near-regular 3-level hierarchical graph we found has more than $15$ nodes.
Therefore, we could either include three-finger combinations or envision the use of the other hand.
We opted for the inclusion of the other hand and kept the combination styles used in our prior work \cite{MentalErrors}, because three-finger combinations would have made the task too demanding.
If we allow such constraints in combination styles, we can have at most $29$ combinations if we do not distinguish which hand presses the spacebar when the spacebar is the only key pressed, without introducing cross-hand combinations.
Cross-hand combinations would also introduce too much complexity to the task.
\textbf{Guideline:} Therefore the range of the number of nodes we aim for is from $15$ to $30$.

\subsection{Degree Homogeneity}
It has been shown that humans are sensitive to local structures \cite{SLInfant} (e.g., degree, and thus transition probabilities)
as well as higher-order structures \cite{SLVisual2, GraphTopoLearning, ProbMotorSeqLearnability} on transition graphs.
To assess whether humans may be sensitive to structures at various global scales,
ensuring regular node degree would eliminate the potential confounding effect due to the variations in transition probabilities.

\begin{figure*}
    \centering
	\includegraphics[width=\textwidth]{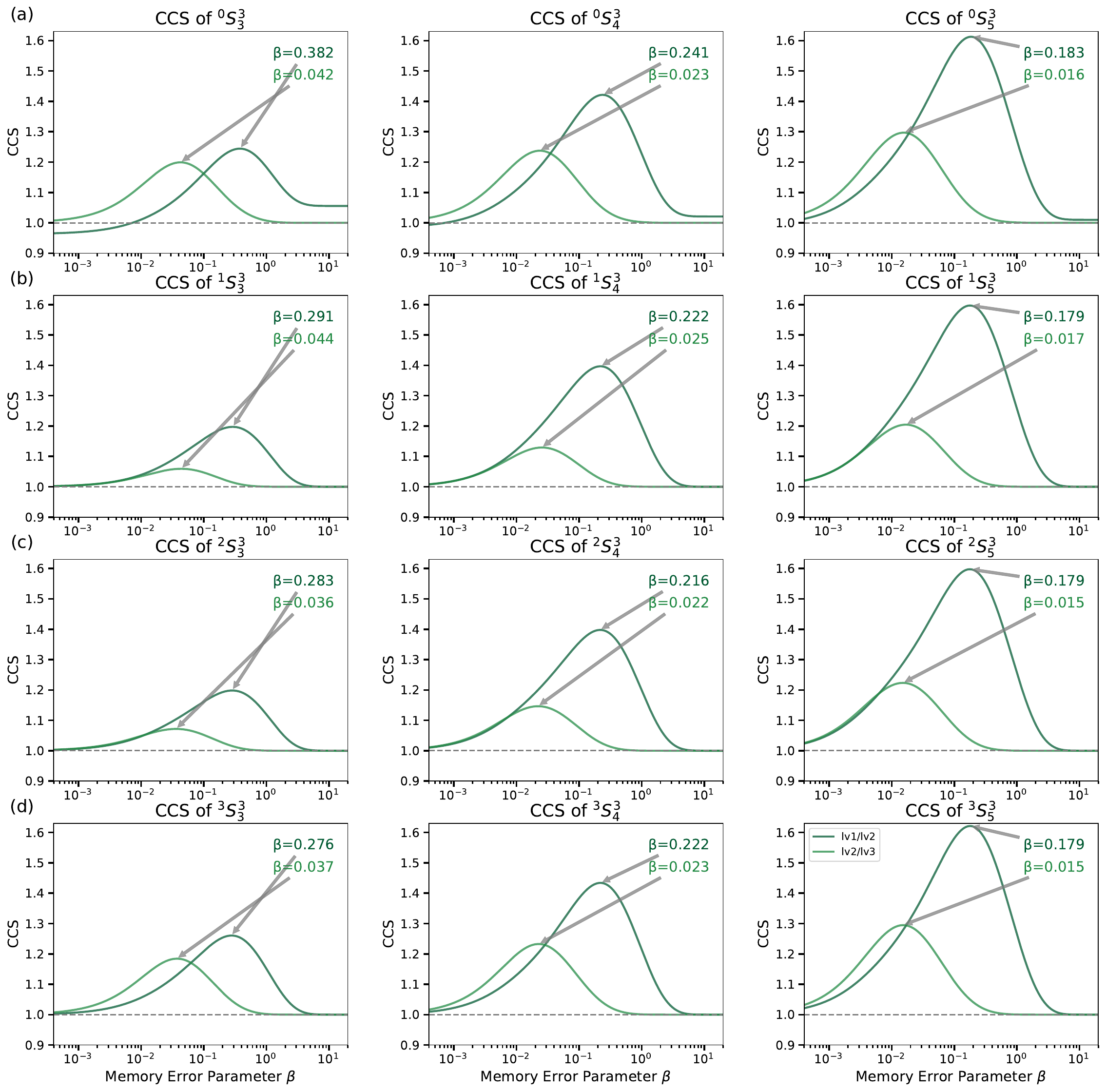}
	\caption{\textbf{The cross-cluster surprisal on power-3 Sierpiński graphs with different bases and regularization types.}
	(a) The CCS for an unregularized 3-level Sierpiński graph across three bases.
	(b) The CCS for a one-node regularized 3-level Sierpiński graph across three bases.
	(c) The CCS for a one-cluster regularized 3-level Sierpiński graph across three bases.
	(d) The CCS for a self-loop regularized 3-level Sierpiński graph across three bases.}
	\label{SI_Fig1.CCS_p}
\end{figure*}

\begin{figure*}
    \centering
	\includegraphics[width=\textwidth]{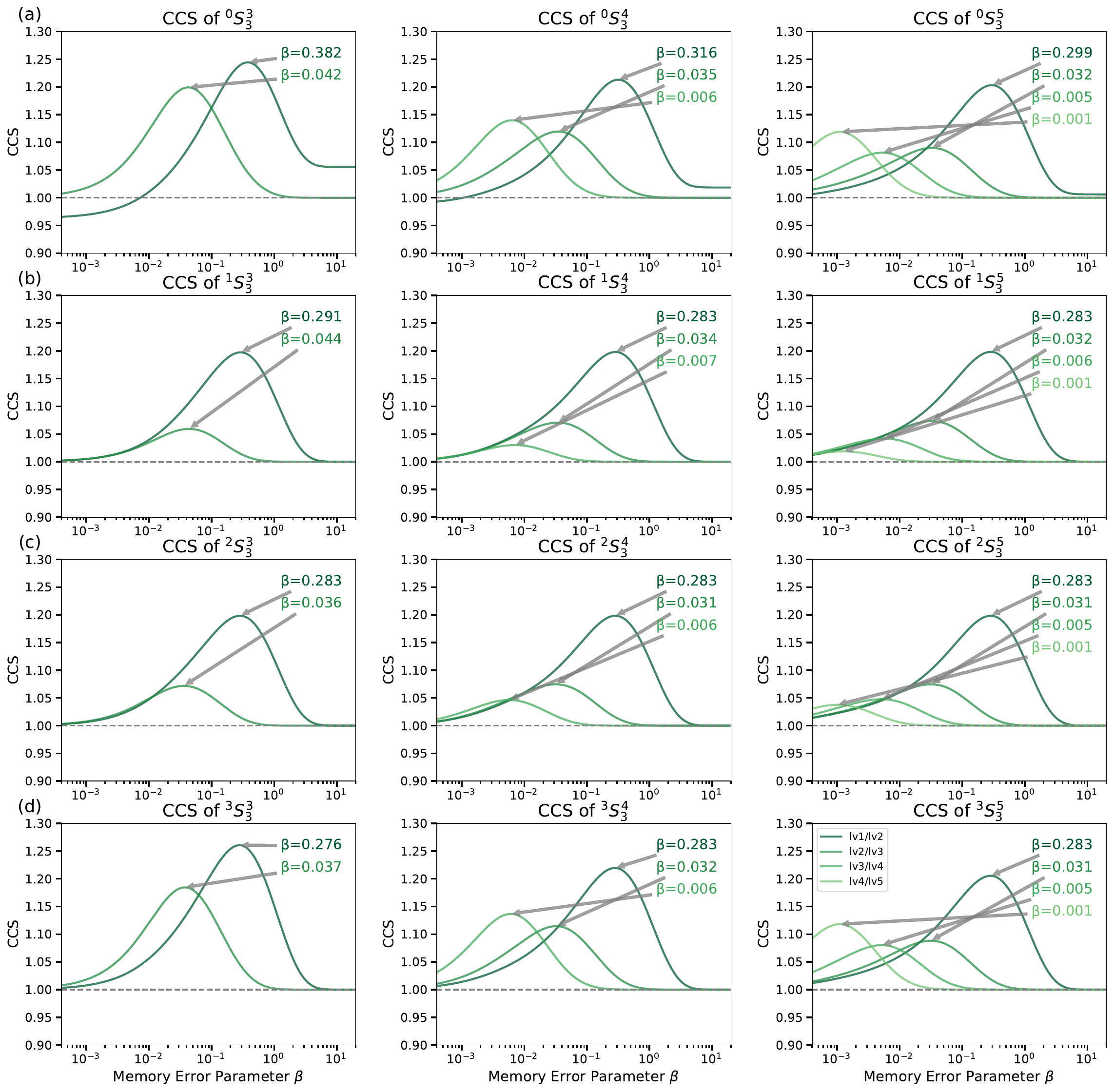}
	\caption{\textbf{The cross-cluster surprisal on base-3 Sierpiński graphs with different levels and regularization types.}
	(a) The CCS for an unregularized base-3 Sierpiński graph across three levels.
	(b) The CCS for a one-node regularized base-3 Sierpiński graph across three levels.
	(c) The CCS for a one-cluster regularized base-3 Sierpiński graph across three levels.
	(d) The CCS for a self-loop regularized base-3 Sierpiński graph across three levels.}
	\label{SI_Fig2.CCS_n}
\end{figure*}

\subsection{Different Types of Graphs in the Sierpiński Graph Family}
We now consider several graph choices and discuss their advantages and disadvantages in relation to the aforementioned guidelines.
We only consider Sierpiński graphs that have three hierarchical levels and have no more than $30$ nodes.
Our range of choices thus narrows down to the sub-family of 3-level, base-3 graphs with different types of regularization.

\subsubsection{Unregularized Sierpiński Graphs}
The $S_3^3$ graph (also denoted as $^{0}S_3^3$) has $27$ nodes, three of which are boundary nodes whose degree is $2$  (which is one less than the degree of the non-boundary nodes).
If we use all $15$ combinations for each hand, we will have $30$ combinations in total, and then will need to remove three of them at random.
We could consider sharing the thumb key-presses across two hands; thus, we would have only $28$ combinations, and would only need to remove one at random.
Apart from the slight key-press complications, the degree distribution breaks the homogeneity assumption, making the graph nearly but not exactly regular, which would complicate the analysis.

\subsubsection{Regularized Sierpiński Graphs}
In the original definition of a Sierpiński graph $S_{p}^{n}$, we observe that there are exactly $p$ boundary nodes $\{0^n,...,(p-1)^n\}$, each of which only has a degree of $p-1$ (which is one less than the degree of non-boundary nodes).
To make the graph regular, which would simplify the analysis, we employ several different regularization methods to ensure that the Sierpiński graph is exactly regular (see \cite{Survey} for an extensive appraisal of regularization methods and other variants of the Sierpiński family).

We denote:
(i) ${}^{0}S_p^n$ (equivalently, $S_p^n$) to be the unregularized Sierpiński graphs;
(ii) ${}^{1}S_p^n$ to be a family of one-node regularized Sierpiński graphs, defined by adding a node that connects to all boundary nodes in $S_p^n$);
(iii) ${}^{2}S_p^n$ to be a family of one-community regularized Sierpiński graphs, defined by adding a smaller Sierpiński graph $S_p^{n-1}$ whose boundary nodes each connects to the boundary nodes in $S_p^n$);
(iv) ${}^{3}S_p^n$ to be a family of self-loop regularized Sierpiński graphs, defined by adding a self-loop to each boundary node in $S_p^n$).
Of note, we excluded ${}^{2}S_3^3$ from the experimental design and analysis because such a regularization would significantly increase the graph size (by $p^{n-1}=9$), which would exceed the desired graph size of $30$ nodes and would make the learning task too demanding without major changes to the task design.
In addition to the pragmatic concern in the guideline, ${}^{2}S_3^3$ would also break its self-similarity property, such that the number of communities at each level is no longer a constant.
The rest of the regularization types listed above, however, also have drawbacks.
One-node regularization introduces shortcuts between boundary nodes and to a lesser extent, their neighbors, changing the average distance between said nodes, and thus affecting the expected count per node in a random walk with a limited size of no more than $3000$.
Self-loop regularization ``traps'' the boundary nodes with a probability of $1/p=1/3$.
In prior work that employed modular graphs \cite{ProbMotorSeqLearnability,MentalErrors}, shortcuts or self-loops did not exist, and they could potentially confound the reaction time on the transitions.

Accordingly, we elected ${}^{3}S_3^3$ as the ground truth graph for participants to learn during a probabilistic sequential motor task.
Although, as previously noted, both ${}^{3}S_3^3$ and ${}^{1}S_3^3$ could potentially confound the experiment with edges that emerge as a result of regularization, we opted for ${}^{3}S_3^3$ because, while satisfying most guidelines, unlike ${}^{1}S_3^3$, it does not introduce a new node that would serve as a shortcut between the communities and complicate the analysis.
Similar to the unregularized case ($^{0}S_3^3$), the overall strength of the CCS does not decrease as fast as that in $^{1}S_3^3$ or $^{2}S_3^3$ when the level increases (Fig. \ref{SI_Fig1.CCS_p}).
A self-loop regularized Sierpiński graph ${}^{3}S_3^3$ has $27$ nodes, each of which corresponds to a unique key combination comprising either one keypress or two keypresses. A slight complication of key combination is that we have to remove one finger combination at random if we share the thumb key-presses across two hands, thus introducing a confounding variable.
However, this confounding variable will be effectively balanced in analysis due to its randomness across participants.

For computations and analyses, we require a definition of edge level for all edges, so that the transitions can be grouped hierarchically.
Thus, we make edge level conventions for the regularized Sierpiński graphs such that, compared to the unregularized $S_p^n$, edges introduced in ${}^{1}S_p^n$ have a level of $n+1$, and edges introduced in ${}^{3}S_p^n$ have an undefined level, or level-$0$.

\section{Coding a Categorical Variable in Linear Regressions}
There is a categorical variable coding scheme called ``forward difference coding'' \cite{forwardcoding2011} that is very similar to the coding theme in this text. We introduce a slightly different coding scheme that we used in the analysis.
Assume that we have a categorical variable $C\in \{C_1, C_2, ..., C_n\}$ ($n \geq 2$) that has $n$ levels.
We can encode $C$ by creating $n$ (or $n-1$, if we use one level as the reference) dummy variables.
Let us consider a simple one dimensional linear model:
\begin{equation}\label{SSLM}
    y = aX + bC + c + \epsilon,
\end{equation}
where $X$ is a continuous independent variable, $c$ is the intercept, and $\epsilon$ is the error term.
One normally would use a series of dummy variables to replace $C$ in Eq. \ref{SSLM}.

\subsection{One-Hot Encoding without Reference}
One way to encode $C$ is to create $n$ dummy variables $\{D_i\}_{1 \leq i \leq n}$,
each of which is binary and has a coefficient $b_i$ ($1 \leq i \leq n$):
\begin{equation}\label{OH_no_Ref}
    y = aX + \sum_{1 \leq i \leq n} {b_i D_i} + c + \epsilon,
\end{equation}
where $D_i \coloneqq \mathds{1}\{C=C_i\}$.
Then for $C=C_i$, Eq. \ref{OH_no_Ref} becomes $y = aX + b_i + c + \epsilon$.
The interpretation of coefficient $b_i$ is thus: how much $C_i$ adds to or subtracts from the outcome $y$ when all other variables are fixed.

\subsection{One-Hot Encoding with a Reference Level}
Similarly, one can encode $C$ by creating $n-1$ dummy variables $\{D_i\}_{1 \leq i \leq n-1}$
relative to reference level $C_n$ (for simplicity we picked $C_n$ as the reference),
each of which is binary and has a coefficient $b_i$ ($1 \leq i \leq n-1$):
\begin{equation}\label{OH_has_Ref}
    y = aX + \sum_{1 \leq i \leq n-1} {b_i D_i} + c + \epsilon,
\end{equation}
where $D_i \coloneqq \mathds{1}\{C=C_i\}$.
Then, for $C=C_i$ ($i \neq n$), Eq. \ref{OH_has_Ref} becomes $y = aX + b_i + c + \epsilon$ (same as one-hot encoding without reference);
however, for $C=C_n$, Eq. \ref{OH_has_Ref} becomes $y = aX + c + \epsilon$.
Coefficient $b_i$ ($1 \leq i \leq n-1$) thus corresponds to
how much $C_i$ adds to or subtracts from the outcome $y$ relative to the outcome when $C=C_n$, when all other variables are fixed.

\subsection{Multi-Hot Encoding with a Reference Level}
Thus far, we have considered encoding the impact of $\{C_i\}_{1 \leq i \leq n}$ on the outcome $y$.
However, we are interested in the CCS, which requires the comparison of two adjacent levels in the categorical variable.
If we define $y_j \coloneqq y | \{C=C_j\}$, then the difference between two adjacent levels is $(y_j - y_{j+1}) \textnormal{ } \forall 1 \leq j \leq n-1$.
To encode $C$ for the differences between two adjacent levels, we create $n-1$ dummy variables $\{D_i\}_{1 \leq i \leq n-1}$
relative to the reference level $C_n$, having selected $C_n$ as reference for simplicity.
Each of the dummy variables is binary and has a coefficient $b_i$ ($1 \leq i \leq n-1$):
\begin{equation}\label{MH_has_Ref}
    y = aX + \sum_{1 \leq i \leq n-1} {b_i D_i(C)} + c + \epsilon,
\end{equation}
where $D_i=D_i(C)$ is effectively a fixed discrete function on the level of the categorical variable $C$.
The specific choice of $n-1$ discrete functions $\{D_j(C)\}$ will be determined shortly.
We now impose a constraint on the interpretation of the coefficient $b_j$ for any $D_j$ such that
said coefficient $b_j$ corresponds to how much $C_j$ adds to or subtracts from the outcome $y$ relative to the outcome when $C=C_{j+1}$, when all other variables are fixed.
This constraint can be written mathematically: $\forall 1 \leq j \leq n-1$,
\begin{equation}\label{constraint}
\begin{split}
    & y_j - y_{j+1} = b_j,\\
    \Leftrightarrow \sum_{1 \leq i \leq n-1} & {b_i \left(D_i(C_j) - D_i(C_{j+1})\right)} = b_j,
\end{split}
\end{equation}
where $D_i(C_j) \coloneqq D_i(C=C_j)$.
Because Eq. \ref{constraint} holds for all permissible choices of $j$ and for any $\{b_i\}_{1 \leq i \leq n-1}$,
we can equate the coefficients of $\{b_i\}_{1 \leq i \leq n-1}$ and simplify Eq. \ref{constraint} to
\begin{equation}\label{constraint_matched}
\begin{cases}
    D_i(C_j) - D_i(C_{j+1}) = 0 & \textnormal{for } i \neq j\\
    D_j(C_j) - D_j(C_{j+1}) = 1
\end{cases}.
\end{equation}
If we restrict $n-1$ discrete functions $\{D_i(C)\}$ to be binary-valued (thus multi-hot) such that $D_i(C) \in \{0,1\}$,
then we can solve for $\{D_i\}_{1 \leq i \leq n-1}$ in Eq. \ref{constraint_matched}:
\begin{equation}\label{uppertriangle}
    D_i(C_j) = \mathds{1}\{j \leq i\} \textnormal{ } \forall 1 \leq j \leq n.
\end{equation}

\subsection{Multi-Hot Encoding in Empirical Data}
In the experimental data, the categorical variable of interest is ``Edgelv'', which has four levels:
``lv0'', ``lv1'', ``lv2'', and ``lv3''.
``lv0'' corresponds to the edges that are introduced through regularization, whose edge level is undefined (zero);
the remaining variables correspond to the edges with respective edge levels.
The three dummy variables are ``lv01'', `lv12'', and `lv23''.
We used the multi-hot encoding (Table \ref{CCS_encoding}) in the linear mixed effects model and we only reported results for the second and third dummy variables ``lv12'' and ``lv23'', because they correspond to the difference definition (as opposed to the ratio definition) of the CCS at level $1$ and level $2$, respectively.

\begin{table}
\setlength{\tabcolsep}{8pt}
\centering
\caption{\textbf{Categorical variable ``Edgelv'' to multi-hot encoded dummy variable conversion.}
The levels with one number in ``Edgelv'' variables correspond to the edge levels,
whereas the levels with two numbers in dummy variables correspond to the difference between two adjacent edge levels.}
\begin{tabular}{lccc}
& \multicolumn{3}{c}{Dummy Variables} \\
Edgelv & lv01 & lv12 & lv23 \\
\hline
self-loop & 1 & 1 & 1 \\
lv1 & 0 & 1 & 1 \\
lv2 & 0 & 0 & 1 \\
lv3 & 0 & 0 & 0 \\

\end{tabular}
\label{CCS_encoding}
\end{table}

\bibliography{main10.5}